\documentclass[11pt,letterpaper]{article}
\usepackage{fullpage}
\usepackage{amsmath}
\numberwithin{equation}{section}
\usepackage{amssymb}
\usepackage{graphicx}
\usepackage{mathrsfs}
\usepackage{cite}
\usepackage{setspace}
\usepackage[dvipsnames]{xcolor}
\usepackage{hyperref}
\hypersetup{
	colorlinks=true,
	linkcolor=blue,
	citecolor=magenta,
	linktoc=page
}
\usepackage{caption}
\ifdefined\pdfsuppresswarningpagegroup
\fi
\begin{document}
	
	
	\vspace{.4cm}
	\begin{center}{\Large \textbf{
				Holographic Entanglement after a Global Quench in \\
				Noncommutative Yang–Mills Theory
	}}\end{center}
	
	\vspace*{15mm}
	\vspace*{1mm}

	\centerline{\bf Elahe Hassanpour, Komeil Babaei Velni and M. Reza Mohammadi Mozaffar}
	
	\bigskip
	
	\centerline{\it Department of Physics, University of Guilan, P.O. Box 41335-1914, Rasht, Iran}
	
	\bigskip
	
	\begin{center}
		{\small E-mails:
			\nolinkurl{elahe_hasanpour@webmail.guilan.ac.ir},
			\nolinkurl{babaeivelni@guilan.ac.ir},
			\nolinkurl{mmohammadi@guilan.ac.ir}}
	\end{center}
	
	\bigskip
	\bigskip
	
	\begin{abstract}
The time evolution of holographic entanglement entropy and mixed-state correlations following a global quench in four-dimensional \(\mathcal N=4\) noncommutative Yang--Mills theory is investigated using a Vaidya geometry. For strip-shaped regions extended along a noncommutative direction, two distinct early-time growth regimes emerge, depending on the relation between the shell-crossing and noncommutative scales: logarithmic growth, with a coefficient proportional to the injected energy density and the square of the noncommutative scale, followed by quadratic growth. For sufficiently wide strips, an intermediate linear-growth regime appears, governed by an entanglement velocity that increases with noncommutativity. The approach to equilibrium is also examined, revealing the role of the butterfly velocity in the continuous evolution of extremal surfaces, while competing branches can modify the physical saturation time. For two disjoint strips, the leading width-independent logarithmic contribution cancels in the mutual information, whereas the entanglement wedge cross section exhibits an intermediate linear-growth regime governed by the same entanglement velocity, provided the connected entanglement wedge remains dominant.
\end{abstract}
	
	\newpage
	
	\tableofcontents

\section{Introduction}
\label{sec:introduction}

Entanglement dynamics following a global quantum quench provides a useful probe of nonequilibrium phenomena in quantum many-body systems \cite{Calabrese:2005in,Calabrese:2009qy}. In integrable theories, the spreading of entanglement can often be understood in terms of entangled quasiparticle pairs \cite{Alba:2017lpnas}. This picture, however, does not generally capture the dynamics of strongly coupled systems, which can instead be studied using the gauge/gravity correspondence \cite{Maldacena:1997re,Gubser:1998bc,Witten:1998qj}. In this setting, the time evolution of entanglement entropy admits a complementary geometric description. In particular, for time-dependent states, the Hubeny--Rangamani--Takayanagi (HRT) prescription relates the entanglement entropy of a boundary region to the area of a bulk extremal surface \cite{Ryu:2006bv,Hubeny:2007xt}. A homogeneous global quench is commonly modeled holographically by an infalling null shell in a Vaidya geometry, which interpolates between an initial state and a final black-brane background \cite{Abajo-Arrastia:2010rvs,Albash:2010mv,Balasubramanian:2010ce,Balasubramanian:2011ur,Keranen:2015fqa}. In holographic theories, this setup typically gives rise to early-time quadratic growth, an intermediate linear-growth regime, and eventual saturation for sufficiently large regions \cite{Liu:2013iza,Liu:2013qca,Li:2013sia,Alishahiha:2014cwa,Fonda:2014ula}. The linear regime is associated with extremal surfaces approaching a critical configuration behind the final horizon and is characterized by the entanglement velocity.

In this work, we investigate how these features are modified by spatial noncommutativity. We consider four-dimensional $\mathcal{N}=4$ noncommutative Yang--Mills (NCYM) theory, with noncommutativity in the $(x_2,x_3)$ plane,
$[x^2,x^3]=i\theta$. Such theories arise from the worldvolume dynamics of D-branes in a background NS--NS $B$-field \cite{Seiberg:1999vs} and exhibit nonlocality and ultraviolet/infrared (UV/IR) mixing \cite{Minwalla:1999px,Douglas:2001ba,Szabo:2001kg}. At large $N$ and strong 't Hooft coupling, their holographic description is provided by the noncommutative D3-brane geometry \cite{Hashimoto:1999ut,Maldacena:1999mh}. The geometry is anisotropic and, in the noncommutative directions, is not asymptotically AdS. Consequently, the entanglement structure depends on the orientation of the boundary region. In particular, for a strip whose finite direction lies along a noncommutative direction, the competing extremal-surface configurations lead to a cutoff-dependent characteristic length scale associated with UV/IR mixing \cite{Barbon:2008ut,Fischler:2013gsa,Karczmarek:2013xxa,Nakajima:2020ncym}. Noncommutative effects on entanglement dynamics and holographic thermalization have also been studied in \cite{Fischler:2018kwt}.

We extend these investigations by studying the time evolution of entanglement entropy and mixed-state correlations in a noncommutative Vaidya geometry. We focus on strips whose finite direction is along $x_2$, and model the quench by an infalling null shell that interpolates between zero- and finite-temperature NCYM backgrounds, keeping the noncommutative scale fixed. Because of the non-AdS asymptotics in the noncommutative directions, we work with a finite radial cutoff and account for the competing extremal-surface branches that arise in this geometry. Our analysis addresses the early-time and intermediate-time scaling of the entanglement entropy, the critical surfaces responsible for linear growth, and the approach to thermal equilibrium. We also examine the resulting dynamics of holographic mutual information and the entanglement wedge cross section (EWCS) for two disjoint strips.

Our results reveal several characteristic effects of noncommutativity. First, the near-boundary behavior of the noncommutative geometry changes the early-time growth of the entanglement entropy: in the regime where the shell-crossing point satisfies $z_c\ll a$, the leading time dependence is logarithmic, with a coefficient proportional to the injected energy density and to $a^2$. When the shell-crossing point lies beyond the noncommutative scale, $a\ll z_c$, the leading behavior instead becomes quadratic. Thus, the early-time scaling depends on the relation between the time-dependent shell-crossing scale and the noncommutative scale, rather than being universally quadratic.

For sufficiently wide strips, we identify a returning branch of shell-crossing extremal surfaces that approaches a critical configuration in the final black-brane geometry. This branch gives rise to an intermediate linear-growth regime, with an entanglement velocity that increases monotonically with the noncommutative scale and reduces to the AdS$_5$ value in the commutative limit. At late times, a branch that connects continuously to the equilibrium extremal surface has a large-strip saturation time controlled by the butterfly velocity along the noncommutative direction. However, competing extremal-surface branches can affect the physical saturation time, which must be determined by comparing their areas.

We further show that the leading, width-independent logarithmic contribution to the early-time entanglement entropy cancels in the connected holographic mutual information of two disjoint strips. Its subsequent evolution is governed by the relative saturation times of the three relevant strip entropies and, in the linear-growth regime, by the same entanglement velocity. Finally, the near-critical connected EWCS exhibits linear growth with a coefficient governed by this velocity. This result applies when the connected entanglement wedge is dominant and the corresponding cross section is the relevant minimal candidate.

The paper is organized as follows. In section \ref{sec:setup}, we introduce the noncommutative Vaidya geometry and the HRT setup. Section~\ref{sec:static-hee} examines the static extremal-surface configurations. In section \ref{sec:time-evolution}, we study the time evolution of the entanglement entropy, including its early-time behavior, critical surfaces, linear growth, and saturation. Section~\ref{sec:mixed} is devoted to the dynamics of mutual information and the EWCS. Finally, section \ref{sec:conclusion} contains our conclusions and a discussion of the physical implications of our results, along
with possible directions for future work.

	\section{Setup}
	\label{sec:setup}
We begin by reviewing the noncommutative D3-brane background and its Vaidya extension, followed by the derivation of the HRT equations for a strip whose finite direction lies along a noncommutative direction.

	\subsection{Noncommutative D3-brane background}
	\label{subsec:nc-Vaidya-geometry}
We consider the bosonic sector of ten-dimensional type IIB supergravity, governed by the following action \cite{Sato:2002kv,Bergshoeff:1995as,Bergshoeff:2005ti}
	\begin{equation}
		\begin{aligned}
			S_{\mathrm{IIB}}
			={}&
			\frac{1}{2\kappa_{10}^{2}}
			\int d^{10}x\,\sqrt{-g}\,
			\Bigg[
			e^{-2\Phi}
			\left(
			R^{(10)}
			+4(\partial\Phi)^2
			-\frac{1}{12}H_3^2
			\right)
			-\frac{1}{2}F_1^2
			-\frac{1}{12}\widetilde F_3^2
			-\frac{1}{4\cdot5!}\widetilde F_5^2
			\Bigg]	\\
			&
			+
			\frac{1}{4\kappa_{10}^{2}}
			\int C_4\wedge H_3\wedge F_3.
		\end{aligned}
		\label{eq:setup-IIB-action}
	\end{equation}
Here $R^{(10)}$ denotes the ten-dimensional Ricci scalar, with $2\kappa_{10}^{2}=16\pi G_N^{(10)}$. We work in the axion-free sector, \(C_0=F_1=0\), so that \(\widetilde F_3=F_3\), and impose the self-duality condition \(\widetilde F_5=\star\widetilde F_5\) after varying the action. The field-strength conventions and equations of motion are given in Appendix~\ref{app:nc-vaidya-consistency}. To accommodate the infalling matter in the Vaidya geometry, we introduce an external source in the string-frame Einstein equation, which takes the form
	\begin{equation}
		R_{\mu\nu} + 2\nabla_\mu\nabla_\nu\Phi - \frac{1}{4}H_{\mu\rho\sigma}H_\nu{}^{\rho\sigma} - \frac{e^{2\Phi}}{4}\left( F_{\mu\rho\sigma}F_\nu{}^{\rho\sigma} + \frac{1}{24}\widetilde F_{\mu\alpha\beta\gamma\delta}\widetilde F_\nu{}^{\alpha\beta\gamma\delta} - \frac{1}{6}g_{\mu\nu}F_3^2 \right) = T^{\mathrm{ext}}_{\mu\nu}.
		\label{eq:setup-Einstein-equation}
	\end{equation}
The external source \(T^{\mathrm{ext}}_{\mu\nu}\) vanishes in the static backgrounds. The dilaton profile is \(e^{2\Phi(z)}=g_s^2 h(z)\), where \(g_s\) is the string coupling. The corresponding finite-temperature noncommutative D3-brane metrics in the string frame are given by \cite{Hashimoto:1999ut,Maldacena:1999mh}\footnote{The
		remaining form fields are given in
		Appendix~\ref{app:nc-vaidya-consistency}.}
	\begin{equation}
		ds^2
		=
		\frac{R^2}{z^2}
		\left[
		-f(z)\,dt^2
		+dx_1^2
		+h(z)\left(dx_2^2+dx_3^2\right)
		+\frac{dz^2}{f(z)}
		+z^2\,d\Omega_5^2
		\right],
		\label{eq:setup-static-metric}
	\end{equation}
	where the blackening function and the warp factor in the noncommutative plane are
	\begin{equation}\label{eq:setup-static-blackening}
		f(z)=1-\left(\frac{z}{z_h}\right)^4,
		\qquad
		h(z)=\frac{z^4}{z^4+a^4}.
	\end{equation}
Here \(R\) is the curvature radius, and \(z_h\) denotes the horizon radius. The noncommutative scale \(a\) is related to \(\theta\) through \cite{Hashimoto:1999ut,Maldacena:1999mh}
	\begin{equation}
		a=\lambda^{1/4}\sqrt{\theta},
		\label{eq:setup-noncommutative-scale}
	\end{equation}
	where \(\lambda=g_{\mathrm{YM}}^2N\) is the 't Hooft coupling. In the commutative limit \(a\to0\), the warp factor satisfies \(h(z)\to1\), whereas for finite \(a\), its near-boundary behavior is \(h(z)\simeq z^4/a^4\). Consequently, the closed-string metric is not asymptotically AdS in the noncommutative directions. We impose a finite radial cutoff \(z=\delta\) throughout the analysis \cite{Barbon:2008ut,Fischler:2013gsa,Karczmarek:2013xxa}. The Hawking temperature, thermal entropy density, and energy density are given by \cite{Maldacena:1999mh}
	\begin{equation}
		T=\frac{1}{\pi z_h},
		\qquad
		s_{\mathrm{th}}
		=
		\frac{R^3}{4G_N^{(5)}z_h^3},
		\qquad
		\mathcal{E}
		=
		\frac{3R^3}{16\pi G_N^{(5)}z_h^4}.
		\label{eq:setup-thermodynamics}
	\end{equation}
	In the following, we express the strip entanglement entropy in units of $\widetilde{\mathcal N}=L^2R^3/(4G_N^{(5)})$. The corresponding normalization of the thermodynamic densities is specified by Eq.~\eqref{eq:setup-thermodynamics}.

\subsection{Noncommutative Vaidya geometry}\label{subsec:nc-Vaidya-spacetime}
	
We model the global quench by an infalling null shell interpolating between the zero- and finite-temperature noncommutative backgrounds, while keeping the noncommutative scale \(a\) fixed. In each static region, we introduce an ingoing Eddington–Finkelstein coordinate through
\begin{equation}
		dv=dt-\frac{dz}{f(z)},
		\label{eq:setup-EF-coordinate}
	\end{equation}
	so that the metric becomes
	\begin{equation}
		ds^2
		=
		\frac{R^2}{z^2}
		\left[
		-f(v,z)\,dv^2
		-2\,dv\,dz
		+dx_1^2
		+h(z)\left(dx_2^2+dx_3^2\right)
		+z^2d\Omega_5^2
		\right],
		\label{eq:setup-Vaidya-metric}
	\end{equation}
	where
	\begin{equation}
		f(v,z)
		=
		1-m(v)z^4.
		\label{eq:setup-Vaidya-blackening}
	\end{equation}
The time dependence enters solely through the blackening function,\footnote{The time-dependent background satisfies the type IIB equations with the external source, as shown in Appendix~\ref{app:nc-vaidya-consistency}.} while the dilaton and form fields retain their static profiles. In the thin-shell limit, the Vaidya mass function \(m(v)\) is given by
\begin{equation}
		m(v)
		=
		\frac{\Theta(v)}{z_h^4}.
		\label{eq:setup-thin-shell-blackening}
	\end{equation}
Here \(\Theta(v)\) is the Heaviside step function. The two static regions are joined across the null shell at \(v=0\), with the time coordinate normalized such that \(v=t\) at the asymptotic boundary. For \(v<0\), the geometry reduces to the zero-temperature noncommutative background, with \(m(v)=0\) and \(f=1\). Integrating \eqref{eq:setup-EF-coordinate} gives
\begin{equation}
    v
    =
    t-z.
    \label{eq:setup-vacuum-time}
\end{equation}
For \(v>0\), the geometry describes the finite-temperature noncommutative black brane, with
\begin{equation}
    f(z)=1-g(z),
		\qquad
		g(z)
		\equiv
		\left(\frac{z}{z_h}\right)^4
\end{equation}
In the exterior region \(z<z_h\), the static and Eddington--Finkelstein time coordinates are related by
\begin{equation}
    v
    =
    t-\int_{0}^{z}\frac{d\zeta}{f(\zeta)}.
    \label{eq:setup-black-brane-time}
\end{equation}
Across the ingoing null shell at \(v=0\), the external stress tensor associated with a general mass profile takes the form
\begin{equation}
    T^{\mathrm{ext}}_{\mu\nu}
    =
    \varrho(v,z)k_\mu k_\nu,
    \qquad
    k_\mu\,dx^\mu=dv,
    \qquad
    \varrho(v,z)
    =
    \frac{3}{2}z^3\,\partial_v m(v),
    \label{eq:setup-null-source}
\end{equation}
where \(k^\mu\) is null. Thus, for the thin-shell profile, the only nonvanishing component is
\begin{equation}
    T^{\mathrm{ext}}_{vv}
    =
    \frac{3z^3}{2z_h^4}\,
    \delta(v),
\end{equation}
where \(\delta(v)\) is the Dirac delta distribution. For the numerical analysis, we use the smooth profile
\begin{equation}
    m(v)
    =
    \frac{m_0}{2}
    \left[
        1+\tanh\left(\frac{v}{v_0}\right)
    \right],
    \label{eq:setup-smooth-profile}
\end{equation}
where \(m_0=1/z_h^4\) and \(v_0>0\) controls the shell thickness. The thin-shell profile is recovered in the limit \(v_0\to0\).	
	
	\subsection{Extremal surfaces and holographic entanglement entropy}\label{subsec:strip-extremal-surfaces}
	We consider a strip of width \(\ell\) along the noncommutative direction \(x_2\), with transverse extent \(L\gg\ell\), defined by
\begin{equation}
		-\frac{\ell}{2}
		\leq
		x_2\equiv x
		\leq
		\frac{\ell}{2},
		\qquad
		-\frac{L}{2}
		\leq
		x_1,x_3
		\leq
		\frac{L}{2}.
		\label{eq:setup-strip-region}
	\end{equation}
In the string frame, the holographic entanglement entropy functional includes the dilaton factor \cite{Klebanov:2007ws}
	\begin{equation}
		S
		=
		\frac{1}{4G_N^{(10)}}
		\int_{\Gamma_\ell}
		d^8\sigma\,
		e^{-2\Phi}
		\sqrt{\det g_{\mathrm{ind}}^{(S)}}.
		\label{eq:setup-string-frame-entropy}
	\end{equation}
Parametrizing the HRT hypersurface by \(v=v(x)\) and \(z=z(x)\), the area functional takes the form

\begin{equation}
	S=\widetilde{\mathcal N}
	\int_{-\ell/2}^{\ell/2} dx\,
	\frac{1}{z^3}
	\sqrt{\frac{\mathcal Q}{h(z)}},
	\label{eq:setup-entropy-ten-dimensional}
\end{equation}

where
	\begin{equation}
		\mathcal Q
		=
		h(z)-f(v,z){v'}^2
		-2v'z',
		\label{eq:setup-Q}
	\end{equation}
and $'\equiv\frac{d}{dx}$. Henceforth, unless stated otherwise, we set the overall normalization factor $\widetilde{\mathcal N}=1$ to unity. Denoting the tip of the extremal hypersurface by \((z_t,v_t)\), we impose the following boundary conditions
\begin{align}
		z\left(\frac{\ell}{2}\right)&=\delta,
		&
		v\left(\frac{\ell}{2}\right)&=t,
		&
		z(0)&=z_t,
		&
		v(0)&=v_t,
		&
		z'(0)&=v'(0)=0.
		\label{eq:setup-boundary-conditions}
	\end{align}
Extremizing the entropy functional ~\eqref{eq:setup-entropy-ten-dimensional} yields the equations of motion
	\begin{align}
		z^3\sqrt{\mathcal Qh}\,
		\partial_x
		\left[
		\frac{z'+fv'}{
			z^3\sqrt{\mathcal Qh}
		}
		\right]
		&=
		\frac{1}{2}\,
		\partial_v f\,v'^2,
		\label{eq:setup-eom-v}
		\\
		z^3\sqrt{\mathcal Qh}\,
		\partial_x
		\left[
		\frac{v'}{
			z^3\sqrt{\mathcal Qh}
		}
		\right]
		&=
		\frac{3\mathcal Q}{z}
		+
		\frac{1}{2}\,
		\partial_z f\,v'^2
		+
		\frac{\partial_z h}{2}
		\left(
		\frac{\mathcal Q}{h}-1
		\right).
		\label{eq:setup-eom-z}
	\end{align}
	The absence of explicit \(x\)-dependence in the Lagrangian leads to the first integral
	\begin{equation}
		z^3
		\sqrt{\frac{\mathcal Q}{h(z)}}
		=
		z_t^3.
		\label{eq:setup-first-integral-J}
	\end{equation}
	In each static region, where \(\partial_v f=0\), an additional conserved quantity follows from Eq.~\eqref{eq:setup-eom-v}
	\begin{equation}
		\frac{z'+fv'}{h(z)}
		=
		E.
		\label{eq:setup-first-integral-E}
	\end{equation}
	In the thin-shell limit, \(E\) is conserved separately in the two static regions, with its values related by the shell-matching conditions. For the smooth profile Eq.~\eqref{eq:setup-smooth-profile}, Eqs.~\eqref{eq:setup-eom-v} and \eqref{eq:setup-eom-z} are solved numerically. We now consider the thin-shell solutions and their matching across \(v=0\).
	
\paragraph{(i) Vacuum region (\(v<0\)).}
The tip conditions ~\eqref{eq:setup-boundary-conditions} imply \(E_-=0\). Eqs.~\eqref{eq:setup-first-integral-J} and
	\eqref{eq:setup-first-integral-E} then yield
	\begin{equation}
		v'=-z',
		\qquad
		z'
		=
		-\frac{\sqrt{h(z)}}{z^3}
		\sqrt{z_t^6-z^6}.
		\label{eq:setup-vacuum-segment}
	\end{equation}
	Choosing the negative root in Eq.~\eqref{eq:setup-vacuum-segment} for the branch extending from the tip toward the cutoff, we obtain
	\begin{equation}
		x(z)
		=
		\int_z^{z_t}
		dy\,
		\frac{y^3}{
			\sqrt{h(y)}
			\sqrt{z_t^6-y^6}
		},
		\qquad
		v(z)=v_t+z_t-z.
		\label{eq:setup-vacuum-profile}
	\end{equation}
	If the surface intersects the shell at $z=z_c$, the condition
	$v(z_c)=0$ yields
	\begin{equation}
		z_c=z_t+v_t.
		\label{eq:setup-shell-position}
	\end{equation}
	\paragraph{(ii) Matching at the shell ($v=0$).}
	The embedding remains continuous across the shell. Integrating \eqref{eq:setup-eom-v} and \eqref{eq:setup-eom-z} across \(v=0\) yields the matching conditions
	\begin{equation}
		v'_+=v'_-,\qquad \mathcal Q_+=\mathcal Q_-,\qquad z'_+=\left[1-\frac{g(z_c)}{2}\right]z'_-.
		\label{eq:setup-shell-refraction}
	\end{equation}
	The subscripts $-$ and $+$ denote the vacuum and black-brane sides, respectively. While the first integral ~\eqref{eq:setup-first-integral-J} is preserved across the shell, the conserved quantity ~\eqref{eq:setup-first-integral-E} on the black-brane side becomes
	\begin{equation}
		E_+=\frac{g(z_c)}{2h(z_c)}\,z'_-.
		\label{eq:setup-shell-energy}
	\end{equation}
	\paragraph{(iii) Black-brane region ($v>0$).} 
Combining the first integrals ~\eqref{eq:setup-first-integral-J} and
	\eqref{eq:setup-first-integral-E} yields	
	\begin{align}
		v'=\frac{h(z)E-z'}{f(z)},\;\;\;\;\;z'^2=h(z)f(z)\left(	\frac{z_t^6}{z^6}-1	\right)+h(z)^2E^2\equiv H(z).
		\label{eq:setup-radial-equation}
	\end{align}
For the monotonic branch, \(z'=-\sqrt{H(z)}\), giving
	\begin{equation}
		\frac{dv}{dz}
		=
		-\frac{1}{f(z)}
		\left[
		1+
		\frac{h(z)E}{\sqrt{H(z)}}
		\right].
		\label{eq:setup-dvdz-final}
	\end{equation}
	The strip width, boundary time, and entanglement entropy are then given by
	\begin{align}
		\frac{\ell}{2}
		={}&
		\int_{z_c}^{z_t}
		\frac{dz}{
			\sqrt{h(z)}
			\sqrt{z_t^6/z^6-1}
		}
		+
		\int_{\delta}^{z_c}
		\frac{dz}{\sqrt{H(z)}},
		\label{eq:setup-width-parametric}
		\\
		t
		={}&
		\int_{\delta}^{z_c}
		\frac{dz}{f(z)}
		\left[
		1+
		\frac{h(z)E}{\sqrt{H(z)}}
		\right],
		\label{eq:setup-time-parametric}
		\\
		S={}&2z_t^3
		\Bigg[
		\int_{z_c}^{z_t}
		\frac{dz}{
			z^3\sqrt{h(z)}
			\sqrt{z_t^6-z^6}
		} +
		\int_{\delta}^{z_c}
		\frac{dz}{
			z^6\sqrt{H(z)}
		}
		\Bigg].
		\label{eq:setup-entropy-parametric}
	\end{align}
	
The above equations provide the basis for our time-dependent analysis. Before examining entanglement evolution, we first study the static extremal hypersurfaces in the vacuum and final black-brane geometries.
	
	\section{Preliminaries: Static backgrounds}
	\label{sec:static-hee}
\begin{figure}[t]
		\centering
		\includegraphics[width=0.7\textwidth]
		{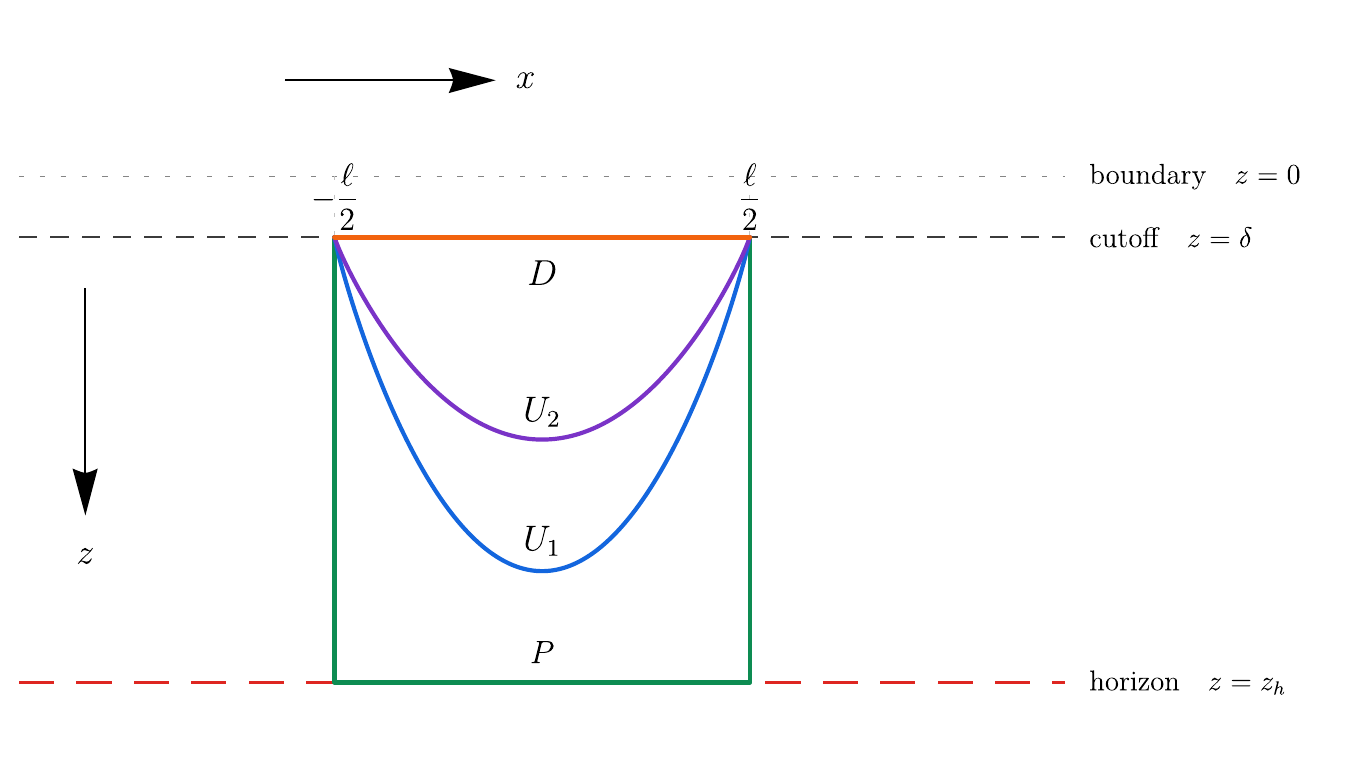}
		\caption{
			Candidate static configurations at finite radial
			cutoff. The surfaces $U_1$ and $U_2$ denote the deeper and
			shallower U-shaped branches, respectively. The configuration
			$D$ lies at the cutoff surface, while the finite-temperature
			configuration $P$ reaches the horizon and includes the horizon
			segment required by homology.
		}
		\label{fig:static-configurations}
	\end{figure}

	We first examine the entanglement entropy in the initial vacuum and final black-brane backgrounds, which provide the reference values for the subsequent time-dependent analysis.
At finite radial cutoff, several candidate RT configurations arise \cite{Fischler:2013gsa}, as illustrated in Figure~\ref{fig:static-configurations}. The U-shaped surfaces comprise a deeper branch \(U_1\) and a shallower branch \(U_2\), while \(D\) denotes the constant configuration at \(z=\delta\). At finite temperature, an additional configuration \(P\) consists of two radial segments reaching the horizon and a connecting horizon segment required by homology. A further U-shaped branch exists near the cutoff, with \(z_t\to\delta^+\) and \(\ell\to0\), distinct from both \(U_2\) and \(D\). We denote the local minimum separating the two bulk U-shaped branches by \(\ell_0\), and the cutoff-dependent scale associated with the competing constant configuration by \(\ell_c\). These definitions apply to both vacuum and thermal backgrounds. Unless stated otherwise, the static numerical results are obtained with \(z_h=1\) and \(\delta=0.05\).
	
	\subsection{Zero-temperature}
	\label{subsec:static-zero-temperature}
	For the vacuum geometry, the static extremal surface lies on a constant-time slice, so the integration constant in
	Eq.~\eqref{eq:setup-first-integral-E} vanishes. Hence, the strip width and entropy become
	\begin{align}
		\ell=2\int_{\delta}^{z_t}	dz\,\frac{z\sqrt{z^4+a^4}}{	\sqrt{z_t^6-z^6}},\;\;\;\;\;\;\;S_{\mathrm{vac}}=2z_t^3\int_{\delta}^{z_t}dz\,\frac{\sqrt{z^4+a^4}}{z^5\sqrt{z_t^6-z^6}}.\label{eq:static-vacuum-entropy}
	\end{align}
	Since the near-cutoff branch extends to arbitrarily small widths, \(\ell_0\) is defined as the local minimum separating the two bulk U-shaped branches,
	\begin{equation}
		\ell_0
		\equiv
		\left.
		\ell(z_t)
		\right|_{
			\partial_{z_t}\ell=0,\,
			\partial_{z_t}^{\,2}\ell>0
		}
		\simeq
		1.6\,a.
		\label{eq:static-vacuum-l0}
	\end{equation}
	The coefficient in the above equation is obtained in the small-cutoff limit \cite{Fischler:2013gsa}. Thus, for \(\ell<\ell_0\), there is no bulk U-shaped solution, although the solutions with \(z_t\simeq\delta\) remain. At finite cutoff, \(U_2\) joins this branch at a local maximum. This maximum moves to arbitrarily large width as \(\delta/a\to0\), and the usual two-branch structure is recovered. Among \(U_1\) and \(U_2\), the deeper branch \(U_1\) has the smaller area \cite{Fischler:2013gsa}.	
	 On the other hand, the constant configuration $D$ has entropy
	\begin{equation}
		S_D(\ell)
		=
		\frac{\ell}{\delta^3}.
		\label{eq:static-cutoff-entropy}
	\end{equation}
	The linear dependence on $\ell$ gives the volume-law behavior of the
	configuration $D$
	\cite{Barbon:2008ut,Fischler:2013gsa,Karczmarek:2013xxa}. The
	near-boundary expansion of the U-shaped surface entropy is
	\begin{equation}
		S_{\mathrm{vac}}^{\mathrm{div}}
		=
		\frac{a^2}{2\delta^4}
		+
		\frac{1}{a^2}
		\log\left(\frac{a}{\delta}\right).
		\label{eq:static-vacuum-divergence}
	\end{equation}
	Comparing this cutoff-dependent expression with Eq.~\eqref{eq:static-cutoff-entropy} gives the following estimate for
	$\ell_c$,
	\begin{equation}
		\ell_c
		=
		\frac{a^2}{2\delta}
		+
		\frac{\delta^3}{a^2}
		\log\left(\frac{a}{\delta}\right)
		\simeq
		\frac{a^2}{2\delta}.
		\label{eq:static-vacuum-effective-cutoff}
	\end{equation}
	Unlike \(\ell_0\), which follows from the branch structure, \(\ell_c\)
	depends on the radial cutoff. The cutoff \(z=\delta\) corresponds
	to a momentum scale of order \(\delta^{-1}\). Because of UV/IR mixing,
	this induces a length scale of order \(a^2/\delta\) along a
	noncommutative direction. This agrees with the leading term in
	Eq.~\eqref{eq:static-vacuum-effective-cutoff}
	\cite{Minwalla:1999px,Fischler:2013gsa,Karczmarek:2013xxa}. We therefore
	restrict our analysis to \(\ell>\ell_c\).
	
	\subsection{Finite-temperature}
	\label{subsec:static-finite-temperature}
	
\begin{figure}[t]
		\centering
		\includegraphics[width=0.5\textwidth]
		{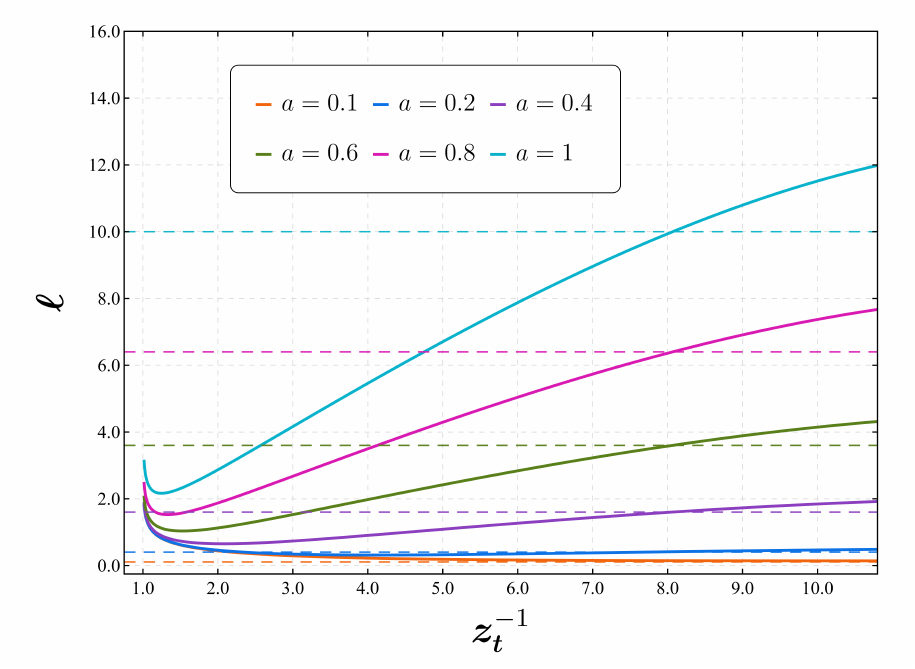}
		\caption{
			Strip width as a function of turning point for several values of the
			noncommutativity parameter. The displayed minima define $\ell_0$.
			The dashed lines
			show the asymptotic estimates of $\ell_c$ from
			Eq.~\eqref{eq:static-thermal-effective-cutoff}. 
		}
		\label{fig:static-thermal-width}
	\end{figure}

	We now consider the final equilibrium state described by the
	finite-temperature black brane. The blackening function is given in
	Eq.~\eqref{eq:setup-static-blackening}, and the turning point satisfies
	$\delta<z_t<z_h$. Again, the minimal surface lies on
	a constant-time slice, and the integration constant in
	Eq.~\eqref{eq:setup-first-integral-E} vanishes. The strip width and entropy are
	\begin{equation}
		\ell=2\int_{\delta}^{z_t}dz\,\frac{z\sqrt{z^4+a^4}}{\sqrt{z_t^6-z^6}\sqrt{1-z^4/z_h^4}},\;\;\;\;	S_{\mathrm{eq}}
		=2z_t^3\int_{\delta}^{z_t}dz\,\frac{\sqrt{z^4+a^4}}{z^5\sqrt{z_t^6-z^6}\sqrt{1-z^4/z_h^4}}.
		\label{eq:static-equilibrium-entropy}
	\end{equation}
	The thermal U-shaped solutions have the same branch structure as in the vacuum geometry.
 At low temperature, this approaches the vacuum result in Eq.~\eqref{eq:static-vacuum-l0}. For $z_h\ll a$, one finds
	\begin{equation}
		\ell_0
		\simeq
		\frac{2a^2}{z_h}
		=
		2\pi a^2T,
		\label{eq:static-thermal-l0-limits}
	\end{equation}
	which shows a linear dependence on temperature \cite{Fischler:2013gsa}. The near-boundary expansion of Eq.~\eqref{eq:static-equilibrium-entropy} gives
	\begin{equation}
		S_{\mathrm{eq}}^{\mathrm{div}}
		=
		\frac{a^2}{2\delta^4}
		+
		\frac{1}{a^2}
		\left(
		1+\frac{a^4}{z_h^4}
		\right)
		\log\left(\frac{a}{\delta}\right).
		\label{eq:static-equilibrium-divergence}
	\end{equation}
	The leading power divergence is independent of temperature, while the logarithmic term receives a thermal correction. Comparing with Eq.~\eqref{eq:static-cutoff-entropy} gives
	\begin{equation}
		\ell_c
		=
		\frac{a^2}{2\delta}
		+
		\frac{\delta^3}{a^2}
		\left(
		1+\frac{a^4}{z_h^4}
		\right)
		\log\left(\frac{a}{\delta}\right)
		\simeq
		\frac{a^2}{2\delta}.
		\label{eq:static-thermal-effective-cutoff}
	\end{equation}
We see that the positive logarithmic correction makes the final-state estimate larger than the vacuum one. As shown in Fig.~\ref{fig:static-thermal-width}, increasing \(a\) shifts the local minimum \(\ell_0\) toward larger strip widths.
	For the parameters used below and for
	sufficiently wide strips, the parallel configuration \(P\) becomes
	relevant only when the U-shaped branch is absent
	\cite{Fischler:2013gsa}.
	At low temperature and small cutoff, where
	\(\ell_c\gg\ell_0\), the U-shaped solutions may still
	exist for widths below \(\ell_c\), although they need
	not be the globally minimal configurations.
	
	\begin{figure}[t]
		\centering
		\begin{minipage}[t]{0.48\textwidth}
			\centering
			\includegraphics[width=\textwidth]
			{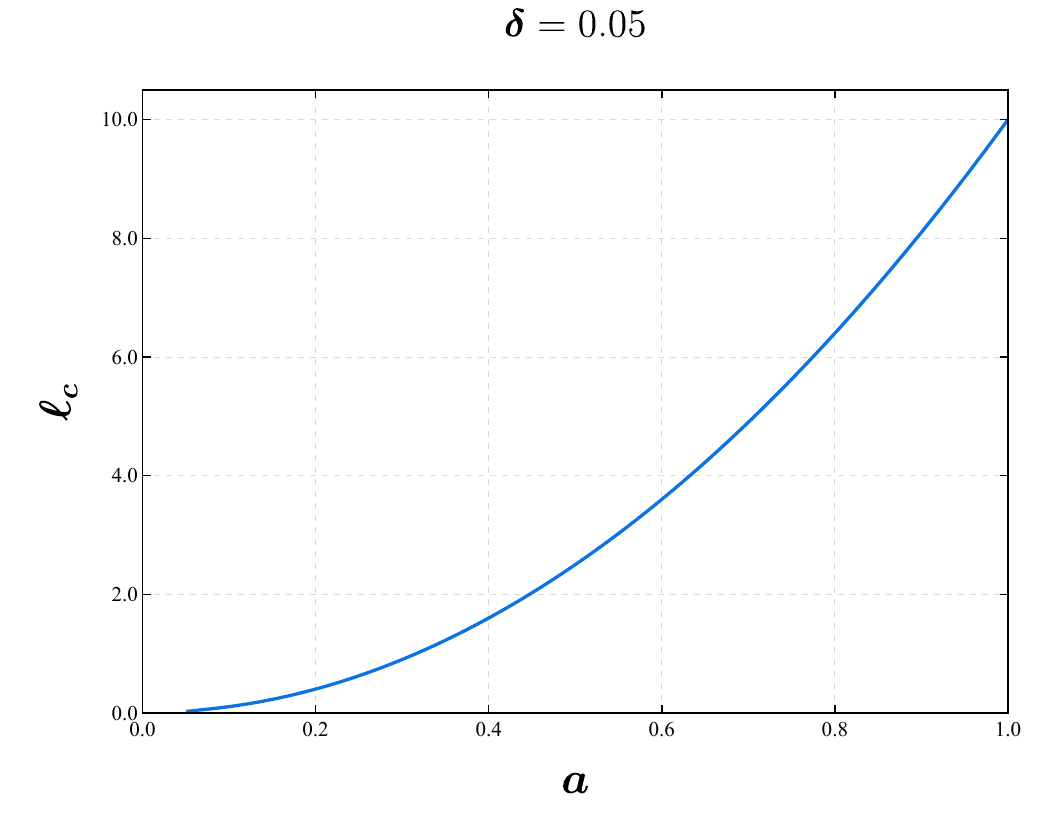}
			\par\smallskip
		\end{minipage}
		\hfill
		\begin{minipage}[t]{0.48\textwidth}
			\centering
			\includegraphics[width=\textwidth]
			{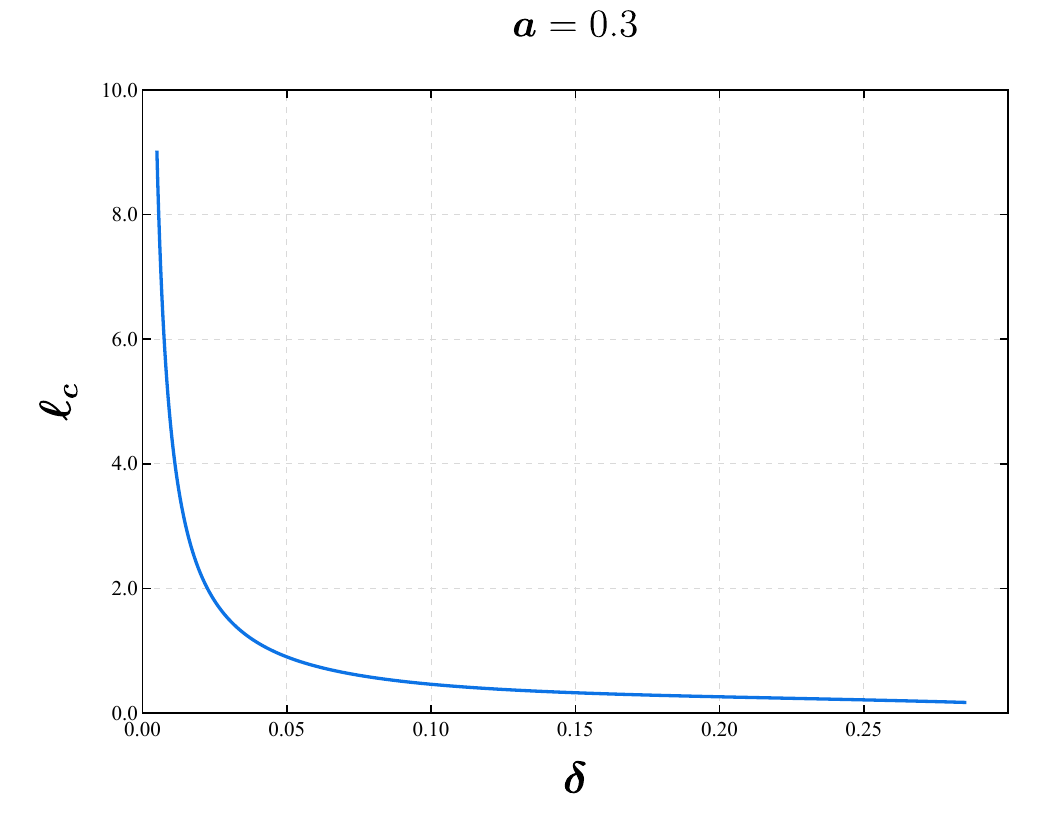}
			\par\smallskip
		\end{minipage}
		\caption{
			Asymptotic final-state estimate of the cutoff scale $\ell_c$ from
			Eq.~\eqref{eq:static-thermal-effective-cutoff}.
			\textit{Left}: As a function of $a$ at fixed cutoff.
			\textit{Right}: As a function of $\delta$ at fixed $a$.
		}
		\label{fig:static-effective-cutoff}
	\end{figure}
	Moreover, Figure~\ref{fig:static-effective-cutoff} shows the final-state estimate of $\ell_c$  increases with $a$. Using the leading estimates in Eqs.~\eqref{eq:static-thermal-l0-limits} and \eqref{eq:static-thermal-effective-cutoff}, the condition $\ell_0>\ell_c$ requires $T\gtrsim1/(4\pi\delta)$. In this range, the temperature is of the same order as the UV cutoff, and the transition between $U_1$ and $P$ depends on the cutoff \cite{Fischler:2013gsa}. We therefore do not consider this range of parameters. Finally, on the $U_1$ branch, the
	equilibrium entropy is extensive for sufficiently wide strips.
	Subtracting the cutoff-dependent terms in
	Eq.~\eqref{eq:static-equilibrium-divergence}, one finds for $\ell T\gg1$
	\begin{equation}
		S_{\mathrm{eq}}(\ell)
		-
		S_{\mathrm{eq}}^{\mathrm{div}}
		=
		\frac{\ell}{z_h^3}
		+
		\mathcal O\left(\ell^0\right),
		\label{eq:static-equilibrium-large-width}
	\end{equation}
	which coincides with the thermal entropy contribution at the leading order.

	\section{Time evolution of entanglement entropy}
\label{sec:time-evolution}

We now investigate the time evolution of
	holographic entanglement entropy following
	the global quench.
 Our analytic treatment
	is based on the thin-shell geometry, whereas
	the numerical calculations employ the smooth
	Vaidya profile introduced in
	Eq.~\eqref{eq:setup-smooth-profile}.
	We define
\begin{equation}
	\Delta S(\ell,t)
	\equiv
	S(\ell,t)-S_{\mathrm{vac}}(\ell),\label{eq:time-entropy-difference}
\end{equation}
where both entropies are evaluated at the same finite radial cutoff. 

	\subsection{Early growth}\label{subsec:early-growth}

	At early times, the HRT surface intersects
	the infalling shell at a radial position
	$z_c$ close to the boundary, satisfying
	$z_c\ll\min\{z_h,z_t\}$.
	The early-time expansion depends on the
	relative magnitude of $z_c$ and the
	noncommutative scale $a$. We therefore
	consider the two limits separately.

	We first consider the regime $\delta\ll z_c\ll\min\{a,z_h,z_t\}$.
The leading time-dependent contribution comes from the portion of the
extremal surface in the black-brane region, $z<z_c$. Expanding
Eqs.~\eqref{eq:setup-width-parametric} and
\eqref{eq:setup-entropy-parametric} at fixed strip width yields
\begin{equation}
	\Delta S(\ell,t)\simeq{}
	\frac{a^2}{z_h^4}\log\left(\frac{z_c}{\delta}\right)
	+\left(
	\frac{1}{8a^2z_h^4}+\frac{7a^2}{48z_h^8}
	\right)z_c^4-\frac{a^2}{12z_h^4z_t^6}z_c^6
	+\cdots.
	\label{eq:early-growth-integrated}
\end{equation}
Expanding Eq.~\eqref{eq:setup-time-parametric} through order $z_c^5$
gives  $
t+\delta=z_c+\frac{7z_c^5}{60z_h^4}+\cdots.
$
Using the large-strip limit of the vacuum width relation, we obtain,
for $\delta\ll t\ll a$ and $t\ll z_h,z_t$,
\begin{equation}
	\Delta S(\ell,t)\simeq{}
	\frac{2\pi A_\Sigma}{3}\mathcal E\,a^2
	\log\left(\frac{t}{\delta}\right)+\left[
	\frac{\pi A_\Sigma}{12a^2}\mathcal E
	+\frac{7\pi^2A_\Sigma^2a^2}{540}\mathcal E^2
	\right]t^4-\frac{32\pi A_\Sigma}{9}
	\left[
	\frac{\sqrt\pi\,\Gamma\!\left(\frac23\right)}
	{\Gamma\!\left(\frac16\right)}
	\right]^6
	\mathcal E\,a^2\left(\frac{t}{\ell}\right)^6
	+\cdots.
	\label{eq:early-growth-energy-density}
\end{equation}
	Here $A_\Sigma=2L^2$ is the area of the entangling surface. The leading growth is logarithmic and independent of the strip width,
while the first explicit width-dependent correction appears at order
$(t/\ell)^6$. To express the logarithmic term as $\log(t/\delta)$,
the additional condition $\delta\ll t$ is required, since
$z_c=t+\delta+\cdots$.
Using Eqs.~\eqref{eq:static-vacuum-divergence} and
\eqref{eq:static-equilibrium-divergence}, the leading logarithmic
result can alternatively be written as
\begin{equation}
	\Delta S(\ell,t)\simeq
	\left(S_{\mathrm{eq}}^{\mathrm{div}}
	-S_{\mathrm{vac}}^{\mathrm{div}}\right)
	+\frac{2\pi A_\Sigma}{3}\mathcal E\,a^2
	\log\left(\frac{t}{a}\right)+\cdots.
	\label{eq:early-growth-fischler-form}
\end{equation}
This form separates the cutoff-dependent static term from the
time-dependent logarithm, which is cutoff independent.
	
	We next consider the regime $\delta\ll a\ll z_c\ll\min\{z_h,z_t\}$.
Expanding Eqs.~\eqref{eq:setup-width-parametric} and
\eqref{eq:setup-entropy-parametric} at fixed strip width,
we obtain the following time-dependent contributions,
omitting time-independent terms:
	\begin{equation}
		\Delta S(\ell,t)\simeq{}
		\frac{z_c^2}{2z_h^4}
	-\frac{a^4}{4z_h^4z_c^2}
		+\frac{z_c^6}{16z_h^8}
		-\frac{z_c^8}{16z_h^4z_t^6}
		+\cdots.
		\label{eq:early-growth-quadratic-integrated}
	\end{equation}
	At fixed strip width, the turning point receives a correction
	$z_t\to z_t+\Delta z_t$, with $\Delta z_t=O(z_c^8)$, as determined by
	Eq.~\eqref{eq:setup-width-parametric}.  Expanding
	Eq.~\eqref{eq:setup-time-parametric} gives
	\begin{equation}
		t+\delta=z_c+\frac{3z_c^5}{40z_h^4}+\cdots.
		\label{eq:early-growth-quadratic-time-expansion}
	\end{equation}
	Combining these results with the large-strip limit of the vacuum width
	relation in Eq.~\eqref{eq:static-vacuum-entropy}, we obtain, for
	$\delta\ll a\ll t\ll\min\{z_h,z_t\}$,
	\begin{equation}
			\Delta S(\ell,t)\simeq{}
			\frac{\pi A_\Sigma}{3}\mathcal E\,t^2
		-\frac{\pi A_\Sigma}{6}\mathcal E\frac{a^4}{t^2}
			-\frac{\pi^2 A_\Sigma^2}{180}\mathcal E^2t^6-\frac{8\pi A_\Sigma}{3}
			\left[
			\frac{\sqrt\pi\,\Gamma\!\left(\frac23\right)}
			{\Gamma\!\left(\frac16\right)}
			\right]^6
			\mathcal E\,t^2\left(\frac{t}{\ell}\right)^6
			+\cdots.
		\label{eq:early-growth-quadratic-energy-density}
	\end{equation}
 The leading time-dependent term is quadratic and proportional to the
	energy density, as in asymptotically AdS--Vaidya geometries
	\cite{Liu:2013iza}. The remaining terms give corrections due to the
	noncommutative scale, higher orders in the black-brane mass, and the
	finite strip width.
In the present geometry, the near-boundary behavior
of the warp factor $h(z)$ modifies the quadratic growth
found in asymptotically AdS--Vaidya geometries and yields
a logarithmic time dependence when $z_c\ll a$. The same near-boundary behavior
	determines the cutoff dependence of the static entanglement entropy in
	NCYM \cite{Barbon:2008ut,Fischler:2013gsa,Karczmarek:2013xxa}.
	Thus, the early-time scaling is not universal and depends on the
	underlying field theory. 
    In the present setup, spatial noncommutativity breaks Lorentz invariance
	and changes the asymptotic geometry. Finally, the logarithmic expansion
	is valid only for $z_c\ll a$. The commutative result therefore cannot
	be obtained by setting $a\to0$ in that expansion; one must take this
	limit in the full equations before expanding near the boundary.

\subsection{Critical surfaces and linear growth} \label{subsec:critical-linear-growth}

 In AdS--Vaidya geometries, the linear growth regime
	is associated with shell-crossing extremal surfaces
	approaching a critical configuration behind the final
	horizon \cite{Liu:2013qca}. 
		For parameter values admitting a boundary anchored branch, a critical shell-crossing
		point $z_c^\ast(z_t)$ exists. As the shell-crossing point approaches
	$z_c^\ast$, the returning surface develops
	an extended portion near the critical radius,
	leading to linear growth of the corresponding entanglement entropy. 

From the refraction condition in
Eq.~\eqref{eq:setup-shell-refraction}, and since
$z'_-<0$, the radial derivative immediately after
crossing the shell becomes positive when
\begin{equation}
	g(z_c)>2,
	\qquad
	z_c>2^{1/4}z_h.
	\label{eq:returning-refraction-condition}
\end{equation}
In this case, the surface moves toward larger
radial depths rather than toward the boundary.
To reach the boundary at $z=\delta<z_c$, it must
reverse its radial direction at a turning point
$z_r$, where $z'=0$ and hence $H(z_r)=0$.
Thus, the above condition is necessary, but not
sufficient, for a returning solution to exist. The turning point satisfies $z_r>z_c$, with
$H(z)>0$ along the black-brane trajectory for
$\delta\leq z<z_r$. For an ordinary turning point,
we also have $H'(z_r)<0$.
At $z_r$, the radial derivative $z'$ changes sign
from positive to negative. Accordingly, we choose
the positive root of Eq.~\eqref{eq:setup-radial-equation}
along the outward-in-depth segment from $z_c$ to
$z_r$, and the negative root along the returning
segment toward the boundary. Hence, taking into account both radial directions of the
black-brane trajectory, the strip width,
boundary time, and entanglement entropy are
given by
\begin{align}
	\frac{\ell}{2}
	={}&
	\int_{z_c}^{z_t}
	\frac{dz}{
		\sqrt{h(z)}
		\sqrt{z_t^6/z^6-1}
	}
	+
	\int_{\delta}^{z_c}
	\frac{dz}{\sqrt{H(z)}}
	+
	2\int_{z_c}^{z_r}
	\frac{dz}{\sqrt{H(z)}},
	\label{eq:setup-returning-width}
	\\
	t
	={}&
	\int_{\delta}^{z_c}
	\frac{dz}{f(z)}
	\left[
	1+\frac{h(z)E}{\sqrt{H(z)}}
	\right]
	+
	2\int_{z_c}^{z_r}
	\frac{h(z)E\,dz}{
		f(z)\sqrt{H(z)}
	},
	\label{eq:setup-returning-time}
	\\
	S
	={}&
	2z_t^3
	\Bigg[
	\int_{z_c}^{z_t}
	\frac{dz}{
		z^3\sqrt{h(z)}
		\sqrt{z_t^6-z^6}
	}
	+
	\int_{\delta}^{z_c}
	\frac{dz}{
		z^6\sqrt{H(z)}
	}
	+
	2\int_{z_c}^{z_r}
	\frac{dz}{
		z^6\sqrt{H(z)}
	}
	\Bigg].
	\label{eq:setup-returning-entropy}
\end{align}
From the matching condition, the integration
constant satisfies $E<0$. At fixed $z_t$, the integration constant
	$E$ depends on the shell-crossing point
	$z_c$ through Eq.~\eqref{eq:setup-shell-energy}.
	For a near-critical hypersurface, $z_r$
	is a simple zero of the radial function $H(z)$
	and lies on the smaller-$z$ side of the critical
	minimum $z_m$, with $z_c<z_r<z_m$.
	As $z_c$ approaches its critical value
	$z_c^\ast(z_t)$ from below, $z_r$ approaches
	$z_m$ from below. At criticality, the two neighboring radial
	roots coincide at $z=z_m$, where the minimum
	becomes a double zero, satisfying
	\begin{equation}
	\left.
	H(z_m)
	\right|_{z_c=z_c^\ast}
	=
	0,
	\qquad
	\left.
	\partial_z H(z)
	\right|_{z=z_m,\,z_c=z_c^\ast}
	=
	0.
	\label{eq:critical-double-zero}
\end{equation}
Combining the above
equations with Eq.~\eqref{eq:setup-radial-equation} and eliminating
\(E\) gives
\begin{equation}
	z_t^6
	=
	\frac{
		z_m^7
		\left[
		\partial_z f(z_m)h(z_m)
		-
		f(z_m)\partial_z h(z_m)
		\right]
	}{
		z_m
		\left[
		\partial_z f(z_m)h(z_m)
		-
		f(z_m)\partial_z h(z_m)
		\right]
		-
		6f(z_m)h(z_m)
	}.
	\label{eq:critical-zt-zm}
\end{equation}
For fixed \(z_t\), this equation determines $z_m$ on the critical branch.
The transition between monotonic and returning
critical configurations occurs when the critical
radius reaches the shell-refraction threshold
$z_s$, defined by $g(z_s)=2$.
At this transition, $z_m=z_c^\ast=z_s$.
Substituting this condition into
Eq.~\eqref{eq:critical-zt-zm}, we find
\begin{equation}
z_t^{(s)}
=
2^{1/4}z_h
\left[
\frac{
	2(4z_h^4+a^4)
}{
	2z_h^4-a^4
}
\right]^{1/6},
\qquad
a^4<2z_h^4.
\label{eq:critical-tip-threshold}
\end{equation}
	For $z_h<z_t<z_t^{(s)}$, the relevant critical
	surface is monotonic, with its characteristic
	radii satisfying $z_m<z_c^\ast<z_s$.
	As $z_t$ increases beyond $z_t^{(s)}$,
	the ordering changes to $z_s<z_c^\ast<z_m$,
	allowing near-critical returning surfaces
	for which $z_s<z_c<z_r<z_m$ and
	$z_c<z_c^\ast$, provided the radial
	turning point is accessible.
	Moreover, in the large-strip limit $z_t\to\infty$,
	the critical equations simplify while
	$z_m$ and $z_c^\ast$ remain finite,
	yielding
\begin{equation}
	z_m
	=
	\left[
	\frac{
		3\left(z_h^4-a^4\right)
		+
		\sqrt{
			9\left(z_h^4-a^4\right)^2
			+
			20a^4z_h^4
		}
	}{2}
	\right]^{1/4},
	\label{eq:critical-zm}
\end{equation}
which satisfies \(z_m>z_h\).
The corresponding critical crossing point
is determined by combining the matching
condition with Eq.~\eqref{eq:critical-double-zero} which gives
\begin{equation}
	z_c^\ast
	=   \Bigg[
	2z_h^4
	\frac{
		\left(z_m^4-z_h^4\right)
		\left(z_m^4+a^4\right)
	}{
		z_m^{10}
	}+
	\sqrt{
		4z_h^8
		\frac{
			\left(z_m^4-z_h^4\right)^2
			\left(z_m^4+a^4\right)^2
		}{
			z_m^{20}
		}
		-
		a^4
	}
	\Bigg]^{1/2},
	\label{eq:critical-zc}
\end{equation}
	Here $z_m$ is determined by Eq.~\eqref{eq:critical-zm}.
		In the large-strip limit, the
		returning critical branch is characterized by
	\[
	0\leq a<2^{1/4}z_h,
	\qquad
	2^{1/4}z_h<z_c^\ast<z_m.
	\]
	At $a=2^{1/4}z_h$, the two critical radii
coincide, $z_c^\ast=z_m=2^{1/4}z_h$,
and the radial interval collapses.
At this endpoint, the leading variation of
$H(z)$ with the crossing position
degenerates in the large-strip limit.
	Although the algebraic expression for
	$z_m$ remains defined beyond this range,
	it does not establish the existence of
	a boundary-reaching profile. 
		\begin{figure}[t]
		\centering
		\includegraphics[width=0.5\textwidth]
		{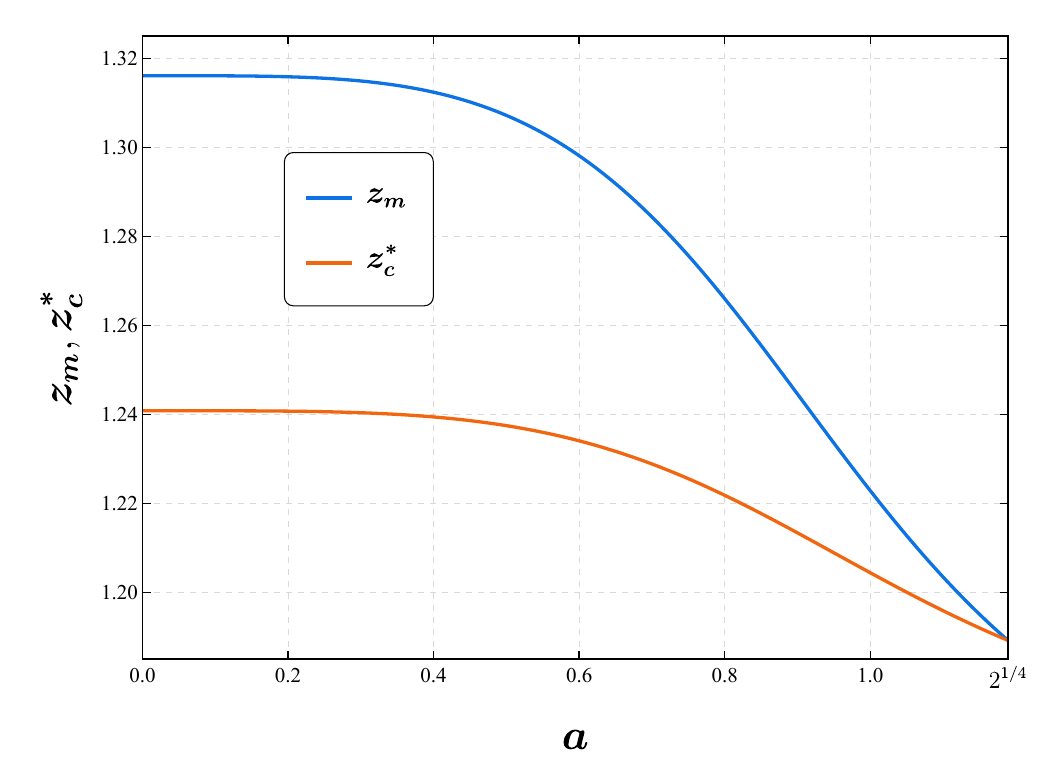}
		\caption{
			Critical radii \(z_m\) and \(z_c^\ast\) as functions of the
			noncommutativity parameter \(a\). The returning branch satisfies
			\(2^{1/4}z_h<z_c^\ast<z_m\) and ends when the two radii coincide.
		}
		\label{fig:critical-scales-a}
	\end{figure}
	In the commutative limit,
	Eqs.~\eqref{eq:critical-zm}
	and \eqref{eq:critical-zc}
	reproduce
	the AdS$_5$--Vaidya result of
	\cite{Liu:2013qca}. Further, both radii decrease with \(a\), and their
	difference also decreases. Near \(a=0\), the change is small because the
	first correction is of order \(a^4\).

A near-critical expansion of the extremal surfaces allows the
linear-growth scaling to be extracted for sufficiently wide
strips in the intermediate-time regime
\(z_h\ll t\ll\ell\). The expansion is parametrized as
\begin{equation}
    z_c
    =
    z_c^\ast(1-\epsilon),
    \qquad
    0<\epsilon\ll1,
    \qquad
    z_c^\ast,z_m\ll z_t.
    \label{eq:linear-near-critical}
\end{equation}
The additional condition
\(z_c^\ast/z_h\ll|\log\epsilon|\)
ensures that the near-critical logarithmic contributions
dominate over the finite terms.
We expand Eqs.~\eqref{eq:setup-returning-width}--\eqref{eq:setup-returning-entropy} near the critical surface, starting with the radial function \(H(z)\)
\begin{equation}
		H(z)
		=H_2(z-z_m)^2+b\,\epsilon+\cdots,\quad
		H_2
		=\frac{1}{2}
		\left.\partial_z^2H\right|_{z=z_m,\,z_c=z_c^\ast},\quad b	=
			-z_c^\ast
			\left.\frac{\partial H}{\partial z_c}	\right|_{z=z_m,\,z_c=z_c^\ast}.
	\label{eq:linear-H-expansion}
\end{equation}
Here the derivative with respect to $z_c$ is
evaluated at fixed $z_t$. For the near-critical branch at fixed
$z_t>z_t^{(s)}$, we have $H_2>0$ and $b<0$.
Since $H(z_r)=0$, the smaller root
approaches $z_m$ according to
	\begin{equation}
		z_r=z_m-\sqrt{-\frac{b}{H_2}}\,\epsilon^{1/2}
		+\mathcal O(\epsilon).
		\label{eq:linear-returning-root}
	\end{equation}
	The logarithmic divergence arises from the returning
	segment upon using Eq.~\eqref{eq:linear-returning-root},
	since
	\begin{equation}
		\int_{z_c}^{z_r}\frac{dz}{\sqrt{H(z)}}
		=-\frac{1}{2\sqrt{H_2}}\log\epsilon
		+\mathcal O(1).
		\label{eq:linear-returning-log}
	\end{equation}
	This segment is traversed twice in the
	parametric expressions, which produces
	logarithmic contributions to the strip width,
	boundary time, and entropy. The remainder here
	is bounded as $\epsilon\to0^+$ at fixed $z_t$.
Using Eq.~\eqref{eq:linear-returning-log}, the leading
	strip-width expression from
	Eq.~\eqref{eq:setup-returning-width} becomes
\begin{equation}
	\frac{\ell}{2}
	\simeq
	c\, z_t
	-
	\frac{1}{\sqrt{H_2}}
	\log\epsilon
	+
	\cdots,
	\label{eq:linear-width-log}
\end{equation}
where
\begin{equation}
	c
	=
	\frac{
		\sqrt{\pi}\,
		\Gamma\left(\frac{2}{3}\right)
	}{
		\Gamma\left(\frac{1}{6}\right)
	}
	+
	\left(\frac{a}{z_t}\right)^4
	\left[
	\frac{1}{2}
	\log\left(\frac{z_t}{a}\right)
	+
	\frac{1}{8}
	+
	\frac{5}{12}\log 2
	\right]
	+
	\mathcal O\left(
	\frac{a^8}{z_t^8}
	\right).
	\label{eq:linear-width-coefficient}
\end{equation}
Further, at the double zero, Eq.~\eqref{eq:setup-radial-equation} fixes
the critical value of \(E\) as follows
\begin{equation}
	E(z_c^\ast)
	=
	-
	\sqrt{
		-\frac{f(z_m)}{h(z_m)}
		\left(
		\frac{z_t^6}{z_m^6}-1
		\right)
	}.
	\label{eq:linear-critical-energy}
\end{equation}
	Expanding Eqs.~\eqref{eq:setup-returning-time}
	and \eqref{eq:setup-returning-entropy}
	near the critical surface, we obtain the leading
	logarithmic contributions to the boundary time
	and the entanglement entropy as
\begin{equation}
	t\simeq-\frac{E(z_c^\ast)h(z_m)}{f(z_m)\sqrt{H_2}}	\log\epsilon+\cdots,\;\;\;\;\;\;\;\;\Delta S(\ell,t)\simeq-
	\frac{
		2z_t^3
	}{
		z_m^6\sqrt{H_2}
	}
	\log\epsilon
	+
	\cdots.
	\label{eq:linear-entropy-log}
\end{equation}
	Eliminating $\log\epsilon$ between the
	width and time relations gives an implicit
	relation for $z_t(\ell,t)$. The leading logarithms in $t$ and $\Delta S$
	originate from the same near-critical returning
	segment. Combining the two relations in
	Eq.~\eqref{eq:linear-entropy-log} and using
	Eq.~\eqref{eq:linear-critical-energy} in the
	limit $z_t\gg z_m$,
	we obtain
\begin{equation}
	\Delta S(\ell,t)
	=
	s_{\mathrm{th}}A_\Sigma
	v_E\,t
	+
	\cdots,
	\label{eq:linear-entropy-linear}
\end{equation}
where we identify
\begin{equation}
	v_E
	=
	\frac{z_h^3}{z_m^3}
	\sqrt{
		-\frac{f(z_m)}{h(z_m)}
	},
	\label{eq:linear-entropy-growth}
\end{equation}
	The linear-growth result is derived for the
	near-critical extremal surface.
	The leading imprint of noncommutativity on the linear-growth regime can be isolated analytically by expanding the entanglement velocity for \(a/z_h\ll1\). This gives
	\begin{equation}
		v_E
		=
		\frac{\sqrt{2}}{3^{3/4}}
		\left[
		1+\frac{1}{6}\left(\frac{a}{z_h}\right)^4
		+\frac{5}{216}\left(\frac{a}{z_h}\right)^8
		+\mathcal{O}\!\left(\left(\frac{a}{z_h}\right)^{12}\right)
		\right],
		\label{eq:vE-small-a}
	\end{equation}
		which shows that noncommutativity enhances the entanglement
		velocity, with the first correction appearing
		at order $(a/z_h)^4$.	This result agrees with the entanglement velocity
	obtained from the shock-wave analysis of thermal
	noncommutative Yang--Mills theory
	\cite{Fischler:2018kwt}. In the commutative limit,
	$v_E\to\sqrt{2}/3^{3/4}$, recovering the
	AdS$_5$ result \cite{Liu:2013iza}.

\subsection{Saturation}\label{subsec:saturation}

The near-critical branch described above does not necessarily
connect continuously to the final equilibrium surface. In
particular, the additional radial turning point \(z_r\) is
specific to the shell-crossing geometry and has no counterpart
in the static equilibrium U-shaped surface. Multiple extremal
branches may therefore coexist near saturation, and the
physical HRT surface is determined by the branch with the
smallest admissible area at each boundary time. In the
late-time limit, the physical extremal surface approaches
the static black-brane configuration. Let $z_b$ denote the turning point of the equilibrium U-shaped
surface with the same strip width. In the parameter range considered in
Sec.~\ref{subsec:static-finite-temperature}, the deeper branch $U_1$ has
the smallest area. If a dynamical branch approaches this surface
continuously, then $E=0$ and $z_c=z_t=z_b$. Equation
\eqref{eq:setup-time-parametric} then gives
\begin{equation}
	t_s^{(\mathrm{cont})}
	=
	\int_{\delta}^{z_b}
	\frac{dz}{f(z)}.
	\label{eq:saturation-continuous-time-integral}
\end{equation}
This is the earliest boundary time at which the equilibrium surface lies
entirely in the final black-brane region. The equilibrium strip width is given by Eq.~\eqref{eq:static-equilibrium-entropy}. In the large entangling region
limit, the turning point of the extremal surface lies close to the
horizon, so that \(z_b=z_h(1-\eta)\) with \(0<\eta\ll1\). Since the final
blackening function has a simple zero at the horizon, the leading terms
are
\begin{equation}
	\frac{\ell}{2}
	=
	-
	\sqrt{
		\frac{
			z_h
		}{
			6|\partial_z f(z_h)|h(z_h)
		}
	}
	\log\eta
	+
	\cdots,
	\qquad
	t_s^{(\mathrm{cont})}
	=
	-
	\frac{\log\eta}{
		|\partial_z f(z_h)|
	}
	+
	\cdots.
	\label{eq:saturation-near-horizon}
\end{equation}
Combining the above results we arrive at
\begin{equation}
	t_s^{(\mathrm{cont})}
	=
	\frac{\ell}{2}
	\sqrt{
		\frac{
			3z_h^4
		}{
			2(z_h^4+a^4)
		}
	}
	+
	\cdots,
	\label{eq:saturation-continuous-time}
\end{equation}
which is a decreasing function of $a$. Another estimate of the saturation timescale can be obtained by
extrapolating the linear-growth expression in
Eq.~\eqref{eq:linear-entropy-linear} to the leading-order
equilibrium entropy. Assuming that the linear-growth regime
persists up to saturation, this yields
\begin{equation}
    t_l
    =
    \frac{\ell}{2v_E}
    +\cdots.
    \label{eq:saturation-linear-time}
\end{equation}
Let us now turn to the approach to the equilibrium
configuration of the extremal surface. Near the saturation regime, both \(z_t\) and \(z_c\) approach the
equilibrium turning point \(z_b<z_h\), so we assume
\begin{equation}
	z_c
	=
	z_t
	\left(
	1-\frac{\epsilon^2}{6}
	\right),
	\qquad
	z_t
	=
	z_b
	\left(
	1+\frac{\sigma}{6}
	\right),
	\qquad
	0<\epsilon,\sigma\ll1.
	\label{eq:saturation-expansion}
\end{equation}
The parameter \(\epsilon\) parametrizes the separation between the shell and the tip, while the fixed-width condition
gives \(\sigma=\mathcal O(\epsilon)\). Substitution into the time
integral and expansion to leading order give
\begin{equation}
	t-t_s^{(\mathrm{cont})}
	=
	\omega\epsilon
	+
	\mathcal O(\epsilon^2),
	\qquad
	\omega
	=
	\frac{g(z_b)}{2\sqrt{h(z_b)}}
	\left[
	\frac{z_b}{
		3f(z_b)^2\mathcal F'(z_b)
	}
	-
	\mathcal J(z_b)
	\right],
	\label{eq:saturation-time-test}
\end{equation}
where
\begin{equation}
	\mathcal F(q)
	=
	q
	\int_{0}^{1}
	\frac{dy}{
		\sqrt{
			f(qy)h(qy)
			\left(y^{-6}-1\right)
		}
	},\qquad
	\mathcal J(q)
	=
	q
	\int_{0}^{1}
	\frac{
		h(qy)\,dy
	}{
		f(qy)
		\sqrt{
			f(qy)h(qy)
			\left(y^{-6}-1\right)
		}
	}.
	\label{eq:saturation-auxiliary-integrals}
\end{equation}
	At fixed strip width, subtracting the equilibrium contribution from
	the shell-crossing entropy yields
	\begin{equation}
		\begin{aligned}
			S(\ell,t)-S_{\mathrm{eq}}(\ell)
			={}&
			2z_t^3
			\left[
			\int_{z_c}^{z_t}
			\frac{dz}
			{z^3\sqrt{h(z)}\sqrt{z_t^6-z^6}}
			+
			\int_{\delta}^{z_c}
			\frac{dz}
			{z^6\sqrt{H(z)}}
			\right]
			\\[1mm]
			&-
			2z_b^3
			\int_{\delta}^{z_b}
			\frac{dz}
			{z^3\sqrt{f(z)h(z)}\sqrt{z_b^6-z^6}} \, .
		\end{aligned}
		\label{eq:entropy-difference-saturation}
	\end{equation}
From Eq. \eqref{eq:saturation-time-test} we see that the sign of $\omega$ determines the behavior near the saturation time. For $\omega<0$,
the HRT surface can approach the equilibrium configuration from earlier times,
whereas for $\omega>0$ it cannot. If $\omega=0$,
higher orders in $\epsilon$ are required
\cite{Liu:2013qca}. For the final static surface, the ingoing time at the tip is
$t-\int_{\delta}^{z_b}dz/f(z)$. Hence, the whole surface lies in the final black-brane region only for
\(t\geq t_s^{(\mathrm{cont})}\). Assuming continuous saturation, we have
\footnote{The derivation is detailed in Appendix~\ref{app:saturation-details}.}
\begin{equation}
	S_{\mathrm{eq}}(\ell)-S(\ell,t)
	\propto
	\epsilon^2
	\propto
	\left[
	t_s^{(\mathrm{cont})}-t
	\right]^2.
	\label{eq:saturation-quadratic}
\end{equation}
For sufficiently large strips, several extremal surfaces coexist
	in the saturation regime, producing a characteristic swallow-tail
	structure in the entropy as a function of boundary time. At saturation,
	the physical HRT surface may switch between competing branches
	when their areas cross and the equilibrium branch becomes globally
	minimal \cite{Albash:2010mv}. Although the transition between competing branches may be
discontinuous at the level of the time derivative, the
entropy itself remains continuous across the transition.
\begin{figure}[t]
	\centering
	\begin{minipage}[t]{0.48\textwidth}
		\centering
		\includegraphics[width=\textwidth]
		{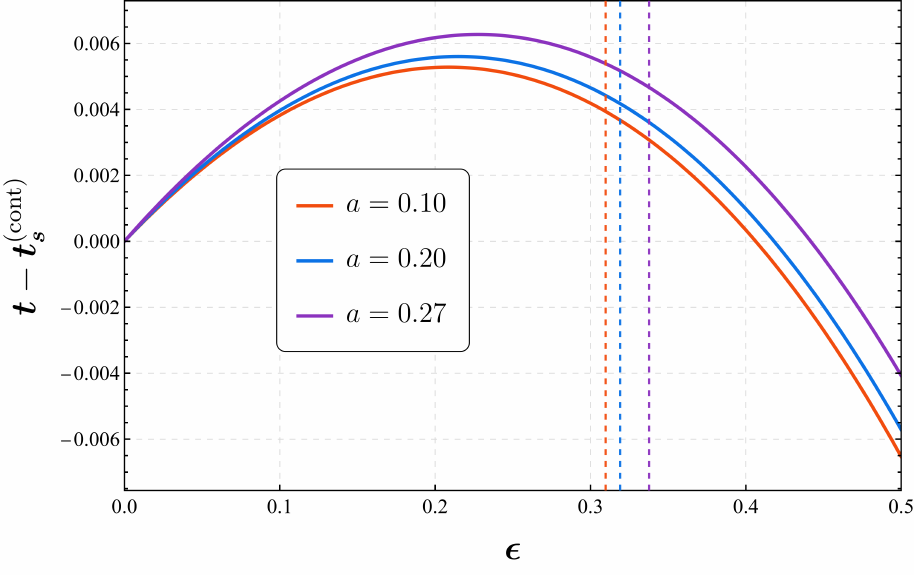}
		\par\smallskip
	\end{minipage}
	\hfill
	\begin{minipage}[t]{0.48\textwidth}
		\centering
		\includegraphics[width=\textwidth]
		{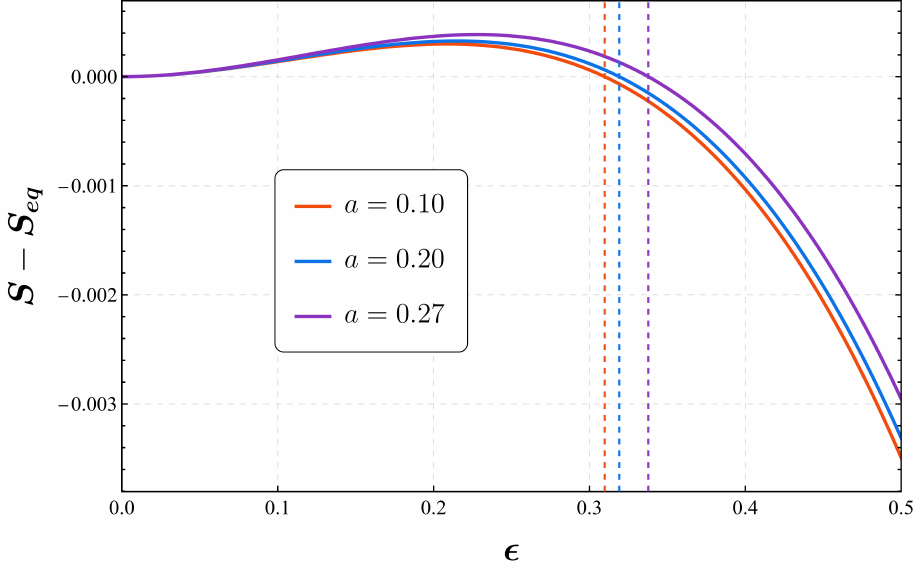}
		\par\smallskip
	\end{minipage}
	\caption{
			Approach of the shell-crossing extremal surfaces to the equilibrium
			surface at fixed strip width.
			\textit{Left}: The time difference
			$t-t_s^{(\mathrm{cont})}$.
			\textit{Right}: The entropy difference
			$S-S_{\mathrm{eq}}$. The dashed lines mark
			$S=S_{\mathrm{eq}}$.
	}
	\label{fig:saturation-ts}
\end{figure}

	The left panel of
	Figure~\ref{fig:saturation-ts} shows the
	dependence of
	$t-t_s^{(\mathrm{cont})}$ on $\epsilon$
	for the shell-crossing extremal surfaces.
	Near $\epsilon=0$, the initial behavior
	is governed by
	Eq.~\eqref{eq:saturation-time-test}.
	The subsequent nonmonotonic behavior
	observed in the numerical curves cannot
	be inferred from this leading-order
	expansion alone.
	As the noncommutativity parameter $a$
	increases, the maximum of the plotted
	time difference shifts toward larger
	values of $\epsilon$.
	The right panel shows the difference
	between the shell-crossing entropy
	and the equilibrium entropy at fixed
	strip width. 
	The intersections indicate candidate
	transitions between the shell-crossing
	and equilibrium configurations.
	Identifying the physical saturation
	point requires selecting the globally
	minimal admissible HRT surface at
	the same boundary time and strip width.

\begin{figure}[t]
	\centering
	\begin{minipage}[t]{0.49\textwidth}
		\centering
		\includegraphics[width=\linewidth]
		{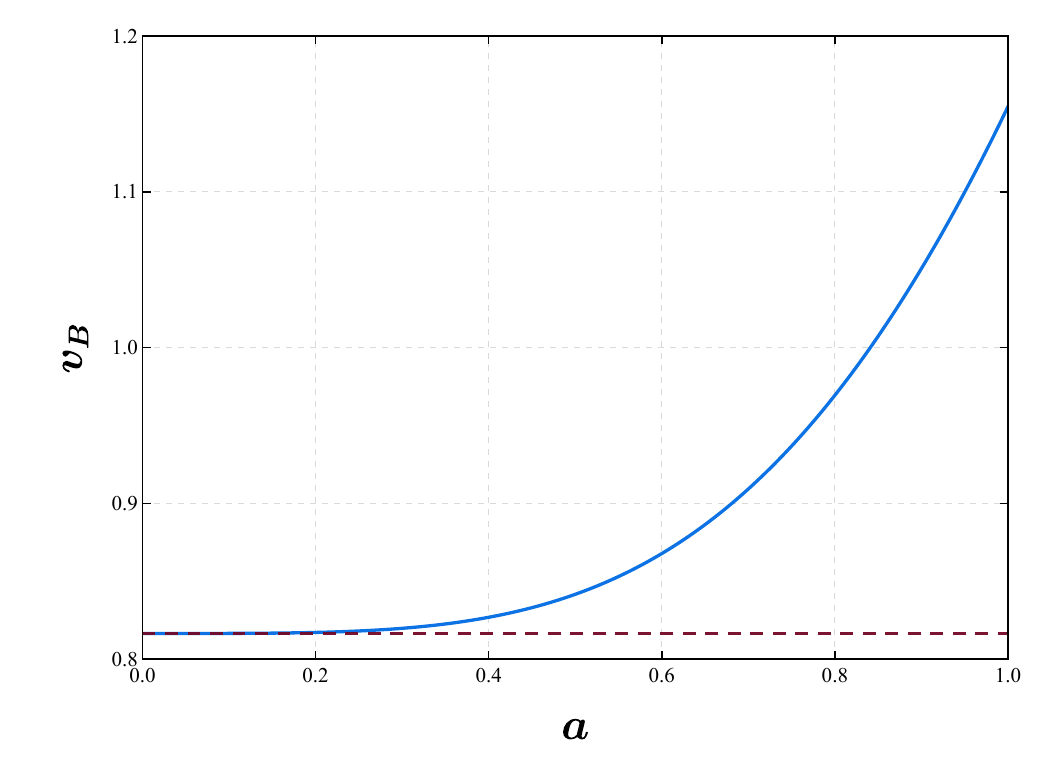}
		\par\smallskip
		
	\end{minipage}
	\hfill
	\begin{minipage}[t]{0.49\textwidth}
		\centering
		\includegraphics[width=\linewidth]
	{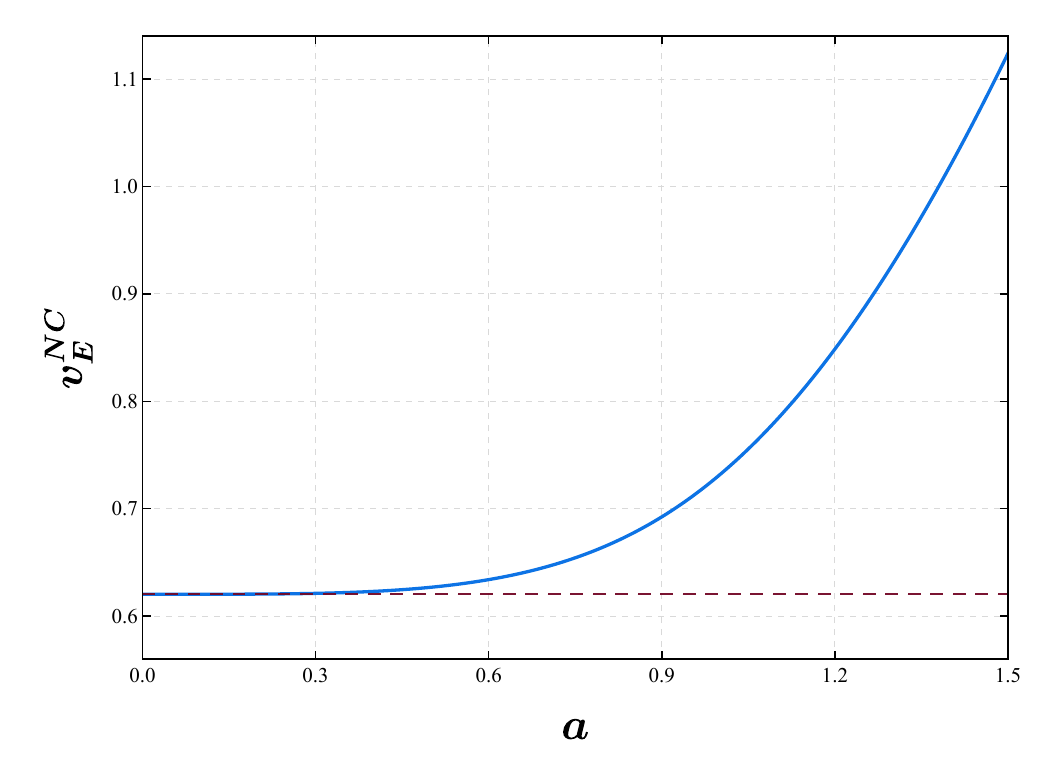	}
		\par\smallskip
		
	\end{minipage}
	\caption{
		\textit{Left}: Butterfly velocity as a function of the
		noncommutative parameter.
		The blue curve shows
		$v_{B,x_2}^2=v_{B,x_3}^2$, while the dashed line
		shows $v_{B,x_1}^2=2/3$. 
		\textit{Right}: Entanglement velocity as a
		function of $a$. The dashed line shows the commutative value.
	}
	\label{fig:linear-rate-velocity}
\end{figure}

To conclude this section, we discuss the relation
between the entanglement velocity at the saturation point and the butterfly velocity $v_B$ which characterizes the
spatial spread of perturbations \cite{Roberts:2016wdl}. The roles of entanglement and butterfly velocities
have also been investigated in anisotropic
black-brane geometries \cite{Jahnke:2018}.
For noncommutative Yang--Mills theory, \(v_B\) was computed in
\cite{Fischler:2018kwt}. Because the geometry is anisotropic, the
result is direction dependent. In our notation,
\begin{equation}
	v_{B,x_1}^2=\frac{2}{3},
	\qquad\qquad
	v_{B,x_2}^2=v_{B,x_3}^2
	=\frac{z_h^4+a^4}{z_h^4}v_{B,x_1}^2.
	\label{eq:butterfly-velocity-nc}
\end{equation}
In terms of \(v_{B,x_2}\) from the above result, Eq.~\eqref{eq:saturation-continuous-time} becomes\begin{equation}
	t_s^{(\mathrm{cont})}
	=
	\frac{\ell}{2v_{B,x_2}}
	+
	\cdots.
	\label{eq:saturation-butterfly-time}
\end{equation}
For sufficiently large \(a/z_h\), \(v_{B,x_2}>1\). Since NCYM is
spatially nonlocal and Lorentz invariance is broken in the
noncommutative directions, this does not imply superluminal signaling in
a local relativistic theory \cite{Fischler:2018kwt}.
	Note that Equation~\eqref{eq:saturation-butterfly-time}
	does not determine the physical saturation
	time when competing HRT branches are present.
	The relation between butterfly propagation
	and entanglement saturation in local
	holographic theories has been discussed
	in \cite{MezeiVanDerSchee:2020}. In the commutative limit, \(v_{B,x_2}\to\sqrt{2/3}\), in agreement with the AdS\(_5\) result \cite{Liu:2013qca}. 
Figure~\ref{fig:linear-rate-velocity} illustrates
the dependence of these velocities on $a$. The left  panel shows that the butterfly velocity along the noncommutative
directions increases monotonically with $a$,
whereas its component along the commutative
direction remains fixed. Further, the right panel shows that the
entanglement velocity increases monotonically
with the noncommutativity parameter $a$.

\subsection{Numerical results}\label{subsec:numerical-entropy-evolution}

\begin{figure}[t]
	\centering
	\begin{minipage}[t]{0.48\textwidth}
		\centering
		\includegraphics[width=\textwidth]
		{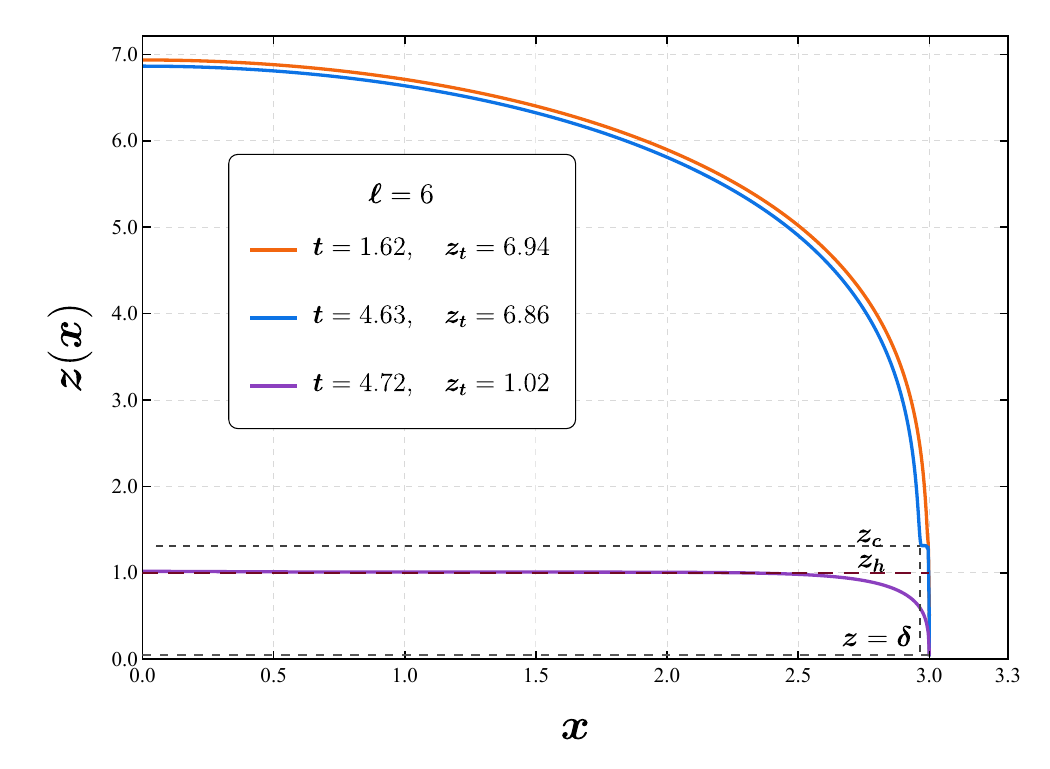}
		\par\smallskip
	\end{minipage}
	\hfill
	\begin{minipage}[t]{0.48\textwidth}
		\centering
		\includegraphics[width=\textwidth]
		{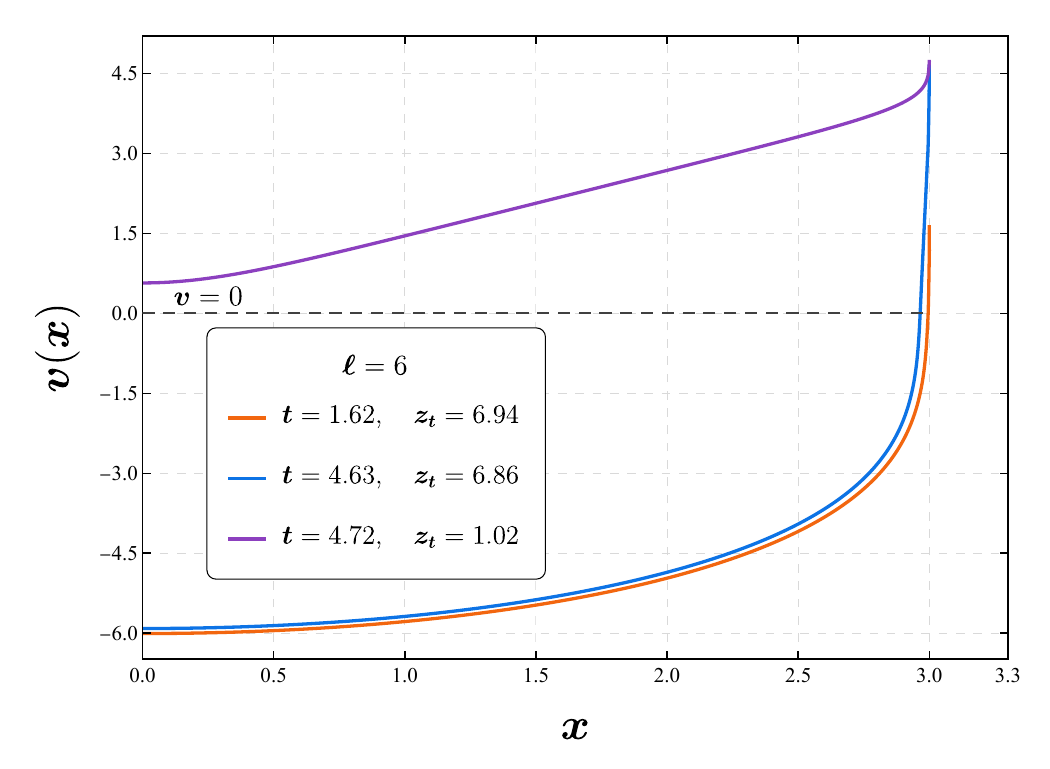}
		\par\smallskip
	\end{minipage}
	\caption{
		Extremal surfaces for the smooth shell at
		$a=0.3$.
		\textit{Left}: The radial profiles $z(x)$; the dashed horizontal
		lines mark the cutoff and the final horizon, and $z_c$ denotes
		the crossing of $v=0$.
		\textit{Right}: The corresponding ingoing-time
		profiles $v(x)$; the dashed line marks $v=0$.
	}
	\label{fig:numerical-HRT-profiles}
\end{figure}

In this subsection we examine the time evolution
	of extremal surfaces using the smooth Vaidya
	profile of Eq.~\eqref{eq:setup-smooth-profile}. Figure~\ref{fig:numerical-HRT-profiles} displays
	the HRT configurations for fixed strip parameters at different boundary
	times. The early-time and intermediate-time surfaces probe both the $v<0$
	and $v>0$ portions of the smooth Vaidya
	geometry. At sufficiently late times, the
	extremal surfaces approach the equilibrium
	configuration of the final black brane.
	These profiles illustrate the geometric
	evolution of the extremal surfaces but do
	not by themselves demonstrate the coexistence
	of competing HRT branches. Establishing such
	competition requires comparing the regulated
	areas of distinct extremal surfaces at the
	same strip width and boundary time, as
	discussed in Sec.~\ref{subsec:saturation}.

\begin{figure}[h!]
	\begin{center}
		\includegraphics[scale=0.36]{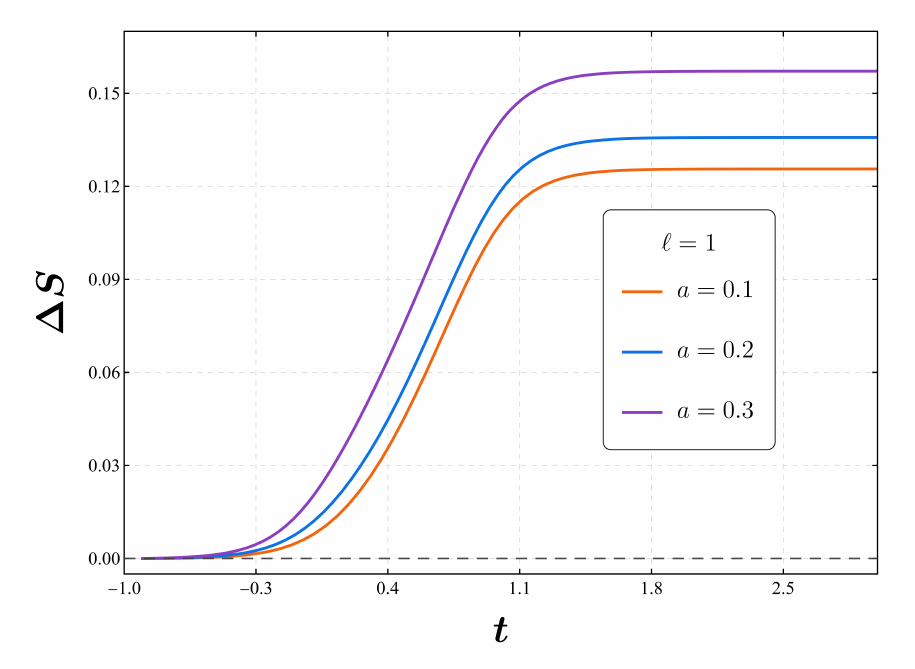}
		\hspace*{-0.4cm}
		\includegraphics[scale=0.36]{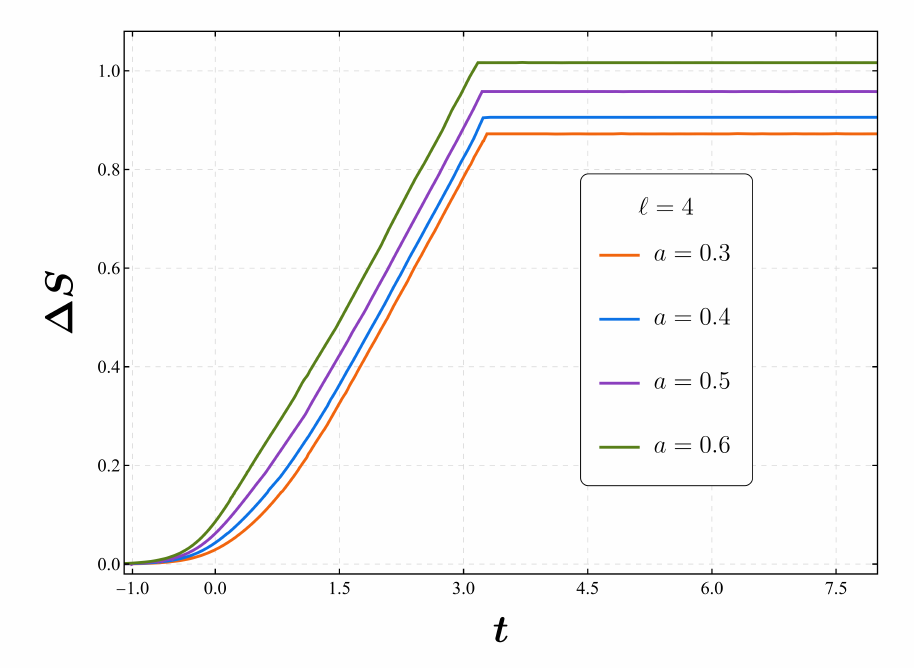}
		\hspace*{-0.4cm}
		\includegraphics[scale=0.36]{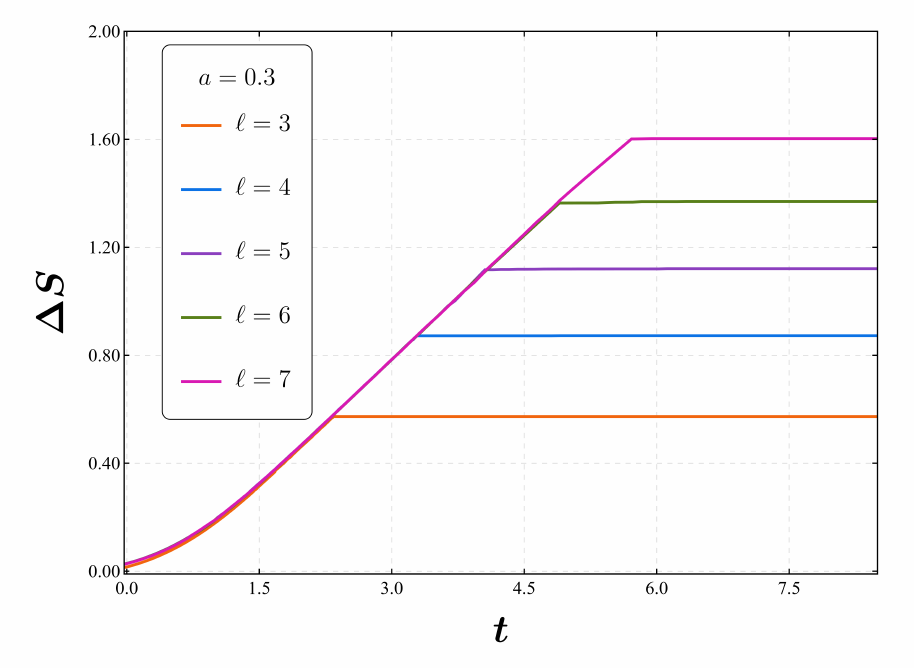}
	\end{center}
	\caption{Time evolution of the entanglement entropy for several values
		of the strip widths and noncommutative scale. The left panel
		corresponds to short regions, while the middle and right panels
		correspond to large regions.}
	\label{fig:numerical-entropy-varying-a}
\end{figure}
Figure~\ref{fig:numerical-entropy-varying-a} shows the time dependence of
the entanglement entropy for several values of the parameters. The entropy increases with time and approaches its equilibrium
value monotonically. As \(a\) increases, \(\Delta S\) grows and the curve
shifts slightly toward earlier times. The latter behavior is consistent
with the dependence of the entanglement velocity on the noncommutativity
scale, as given by Eq.~\eqref{eq:linear-entropy-linear}, and with the
increase of $v_E$ reported in the right panel of
Figure~\ref{fig:linear-rate-velocity}. 
For large regions, the selected HRT branch can
	change at saturation; this need not produce a jump
	in the entropy, although it can produce a kink. In particular,
the right panel shows the evolution for a fixed value of \(a\) in the large \(\ell\) limit. As \(\ell\) increases, the
linear-growth regime becomes more pronounced. The curves nearly coincide until the smaller
strips begin to saturate. Once a strip reaches equilibrium, its entropy
becomes constant, while the wider strips continue along nearly the same
growing curve.
At early times, this behavior is consistent with the
width-independent coefficient in
Eq.~\eqref{eq:early-growth-energy-density}.
In the linear regime, it agrees with
Eq.~\eqref{eq:linear-entropy-linear}, whose leading slope does not depend
on \(\ell\).
	Increasing the strip width therefore extends
	the linear-growth regime and delays saturation
	without changing the leading growth rate.
	The numerical curves mainly display the full evolution of the
	entropy up to saturation. On this scale, a very short initial
	logarithmic regime may be obscured by the subsequent stages
	of growth. Furthermore, the temporal resolution of the
	numerical data may be insufficient to identify such a short
	regime.

\begin{figure}[t]
	\centering
	\begin{minipage}[t]{0.48\textwidth}
		\centering
		\includegraphics[width=\textwidth]
		{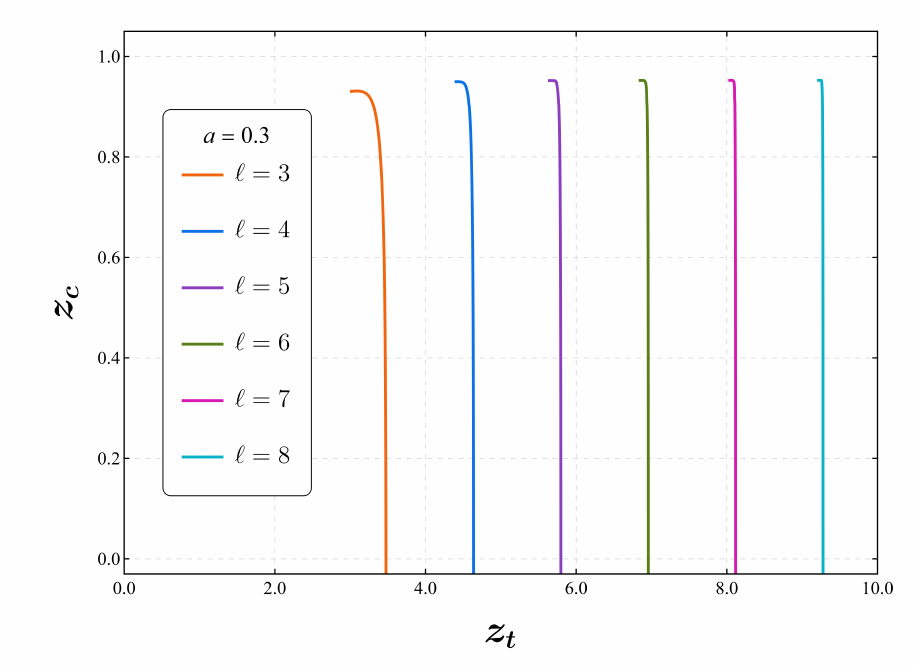}
		\par\smallskip
	\end{minipage}
	\hfill
	\begin{minipage}[t]{0.48\textwidth}
		\centering
		\includegraphics[width=\textwidth]
		{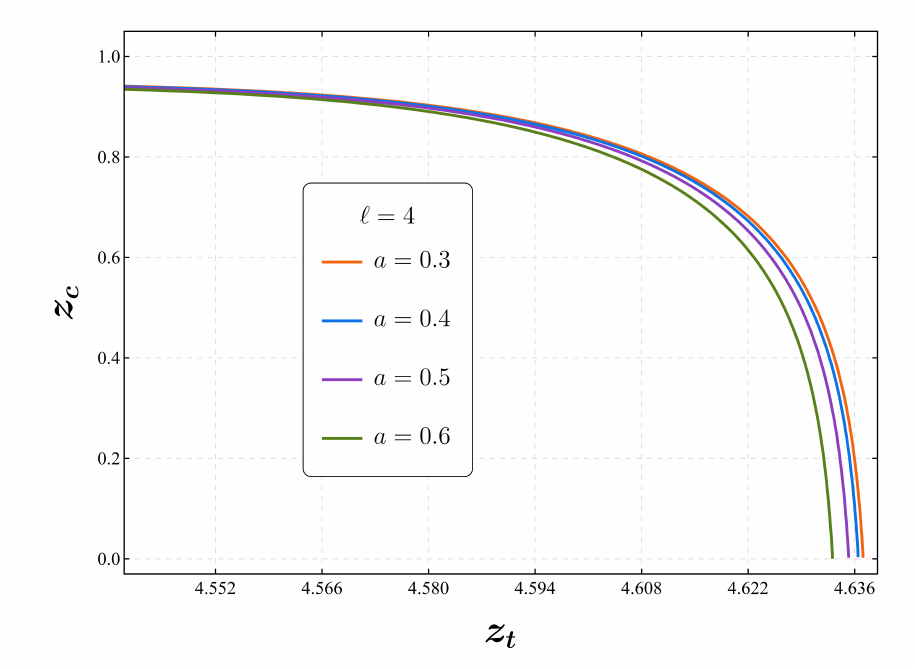}
		\par\smallskip
	\end{minipage}
	\caption{
		\(z_c\) as a function of \(z_t\).
		\textit{Left}: For several values of the strip width.
		\textit{Right}: For several values of the noncommutative scale.
		For wide strips, the shell-crossing positions
			approach the thin-shell critical scale approximately.}
	\label{fig:numerical-zc-zt-branches}
\end{figure}

The left panel of Figure~\ref{fig:numerical-zc-zt-branches} shows the parametric plots of
\((z_c(\ell,t),z_t(\ell,t))\) for several values of \(\ell\). Increasing the strip width shifts the curves toward larger turning points, so that wider strips probe deeper regions of the bulk. For sufficiently wide strips, the shell-crossing
	positions approach a common limiting region, consistent
	with the critical scale of the thin-shell analysis in
	Sec.~\ref{subsec:critical-linear-growth}. At finite strip
	width and nonzero shell thickness, the agreement with
	$z_c^\ast(z_t)$ is approximate. The right panel of Figure~\ref{fig:numerical-zc-zt-branches} illustrates
the same functions for several values of the noncommutative
scale at fixed strip width. At a given \(z_c\), increasing \(a\) shifts
the curve toward slightly smaller \(z_t\). The corresponding extremal
surfaces therefore become slightly shallower as the noncommutative scale
is increased. This effect is small compared with the shift produced by
changing the strip width.

	\section{Mixed-state correlations after the quench}\label{sec:mixed}
	
	Having established the time evolution of the entanglement entropy for a
	single strip, we now turn to mixed-state correlation measures. We focus
	on the mutual information and the entanglement wedge cross section, which
	probe correlations between two disjoint regions.
	The entanglement wedge cross section has been linked
	to several mixed-state information measures,
	including entanglement of purification, reflected
	entropy, and odd entanglement entropy.
	Logarithmic negativity provides another measure
	of mixed-state entanglement \cite{Calabrese:2012ew},
	and holographic proposals involving entanglement-wedge
	cross sections have also been explored for this
	quantity \cite{KudlerFlam:2018qjo}.
	 The setup is the same as before:
	the noncommutative scale \(a\) is fixed, and the time dependence enters
	only through the Vaidya mass function. Since both quantities involve extremal surfaces, the techniques developed in the previous section provide a basis for the analysis below.

	\subsection{Mutual information}\label{sec:hmi}

	The mutual information between two disjoint regions $A$ and $B$, which measures their total correlations, is defined as
	\begin{equation}
		I(A:B)
		=
		S(A)+S(B)-S(A\cup B).
		\label{eq:mutual-information-intro}
	\end{equation}
	In local quantum field theories, the mutual information between
	disjoint regions is ultraviolet finite \cite{Headrick:2010zt}.
	Holographically, it is determined by the connected and disconnected
	RT/HRT configurations. The transition between these configurations
	gives rise to a phase transition in the large-$N$ limit
	\cite{Headrick:2010zt,Fischler:2012uv}. Holographic mutual information in
	static NCYM was studied in \cite{Fischler:2013gsa}.

\begin{figure}[t]
		\centering
		\includegraphics[width=0.82\textwidth]
		{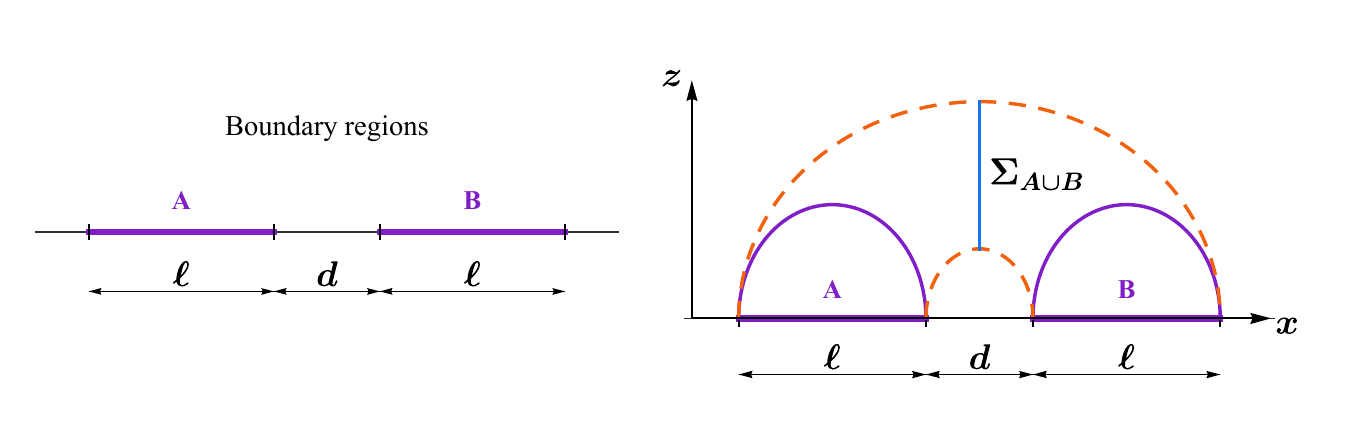}
		\caption{
			Schematic connected and disconnected HRT configurations for the
			two-strip setup. The blue segment denotes the cross section of the
			connected entanglement wedge and is discussed in
			Sec.~\ref{sec:ewcs}.
		}
		\label{fig:hmi-configurations}
	\end{figure}

	We consider two disjoint strips $A$ and $B$ of equal width $\ell$,
	separated by a distance $d$. Their finite directions lie along the
	noncommutative coordinate $x_2$. The two candidate HRT configurations
	are shown in Figure~\ref{fig:hmi-configurations}. We work in the regime $\{d,\ell\}>\ell_c$, where $\ell_c$ is the
	cutoff-dependent scale of the final black-brane geometry defined in
	Eq.~\eqref{eq:static-thermal-effective-cutoff}. 
	 For $A\cup B$, we consider two non-crossing HRT configurations. The
	disconnected configuration consists of two surfaces of width $\ell$.
	The connected configuration is formed by the surfaces of widths $d$
	and $2\ell+d$. Hence,
	\begin{equation}
		S(A\cup B;t)
		=
		\min\left\{
		2S(\ell,t),\,
		S(d,t)+S(2\ell+d,t)
		\right\}.
		\label{eq:hmi-union-entropy}
	\end{equation}
	Using the definition in Eq.~\eqref{eq:mutual-information-intro}, the
	holographic mutual information is therefore
	\begin{equation}
		I(A:B;t)
		=
		\max\left\{
		0,\,
		2S(\ell,t)-S(d,t)-S(2\ell+d,t)
		\right\}.
		\label{eq:hmi}
	\end{equation}
	If the connected configuration is dominant, the mutual information is
	nontrivial. Otherwise, the disconnected configuration is dominant and the
	leading classical contribution to the mutual information vanishes. At the
	transition point, the two configurations are degenerate
	\cite{Headrick:2010zt}. 
	We divide the evolution of the holographic mutual
	information into three parts and discuss each separately.
	
\textbf{Early time.} 
We consider the regime $\ell\gg d$ where the connected HRT configuration
is dominant during the time interval under consideration.
At early times, the coefficient of the leading logarithmic
term in Eq.~\eqref{eq:early-growth-energy-density} is
independent of the strip width. Consequently, this
contribution cancels in the connected combination in
Eq.~\eqref{eq:hmi}. 
The leading time dependence of the mutual information is therefore
determined by subleading width-dependent terms, which
are not fixed by the leading-order result for the
entanglement entropy.

\textbf{Intermediate time.}
In this regime, Eq.~\eqref{eq:linear-entropy-linear} gives the same
leading growth rate \(s_{\mathrm{th}}A_\Sigma v_E\) for each strip. When
all three entropies in Eq.~\eqref{eq:hmi} lie in the common linear
regime, the leading linear terms cancel and the mutual information is
approximately constant. The entropies for regions of length \(d\),
\(\ell\), and \(2\ell+d\) then reach their equilibrium values successively.
After \(S(d,t)\) has saturated while the other two entropies continue to
grow linearly, the slope of the mutual information becomes
\(s_{\mathrm{th}}A_\Sigma v_E\). Once the width-\(\ell\) entropy also
becomes approximately constant while the \(2\ell+d\) entropy continues to
grow, the slope becomes \(-s_{\mathrm{th}}A_\Sigma v_E\). The mutual
information can therefore increase and then decrease, with a maximum
between these two regimes. The crossover times depend on the strip widths, their separation, and the parameters of the quench.

\begin{figure}[t]
		\centering
		\includegraphics[width=0.5\textwidth]{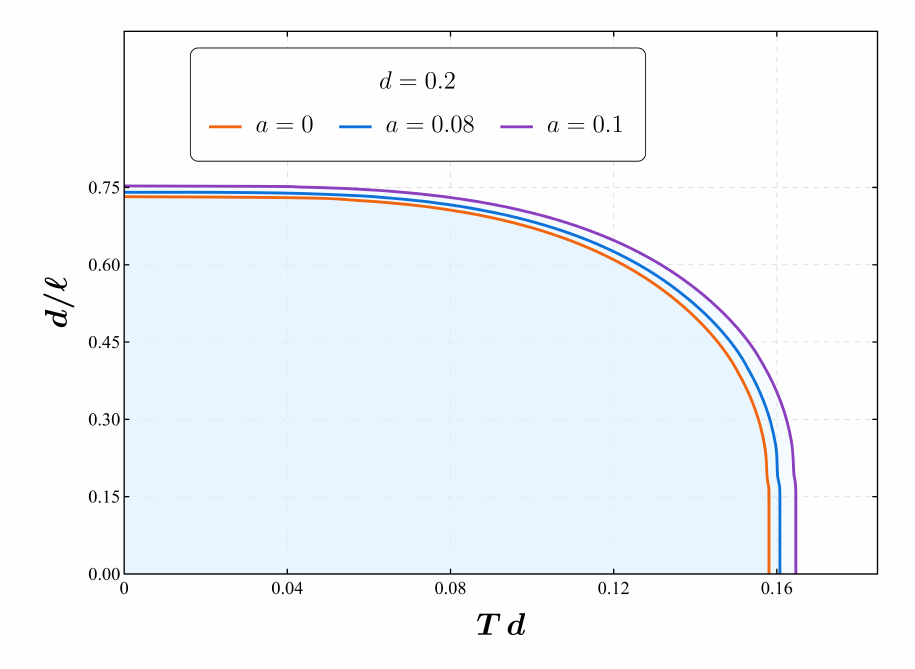}
		\caption{Phase diagram of the holographic mutual information
		in the $(Td,d/\ell)$ plane for fixed $d$. The shaded region corresponds to
			nonzero mutual information.}
		\label{fig:HMI_phase_diagram}
	\end{figure}

\textbf{Late time.}
	At late times, the HRT surfaces approach their final black-brane
	configurations. The equilibrium mutual information follows from
	Eq.~\eqref{eq:hmi} after replacing the strip entropies by the
	corresponding equilibrium values. Here we refer to the leading classical
	contribution described by the HRT areas; bulk correlations of order
	\(N^0\) are not included. 
			To characterize the connected-to-disconnected transition
		in the final equilibrium state, we determine the phase
		transition point from the condition
		\[
		2S_{\mathrm{eq}}(\ell)-S_{\mathrm{eq}}(d)
		-S_{\mathrm{eq}}(2\ell+d)=0.
		\]
		Figure~\ref{fig:HMI_phase_diagram} shows the resulting
		phase diagram in the $(Td,d/\ell)$ plane for fixed
		$d=0.2$ and different values of the noncommutative
		scale $a$. For each curve, the region below the phase
		boundary corresponds to nonzero equilibrium mutual
		information, whereas above it the disconnected
		configuration dominates and the leading holographic
		mutual information vanishes. Increasing the
		noncommutativity parameter shifts the phase boundary
		toward larger values of $Td$ and $d/\ell$, thereby
		enlarging the region of nonzero mutual information
		within the parameter range considered.
	
\begin{figure}[t]
		\centering
		\begin{minipage}[t]{0.49\textwidth}
			\centering
			\includegraphics[width=\textwidth]
			{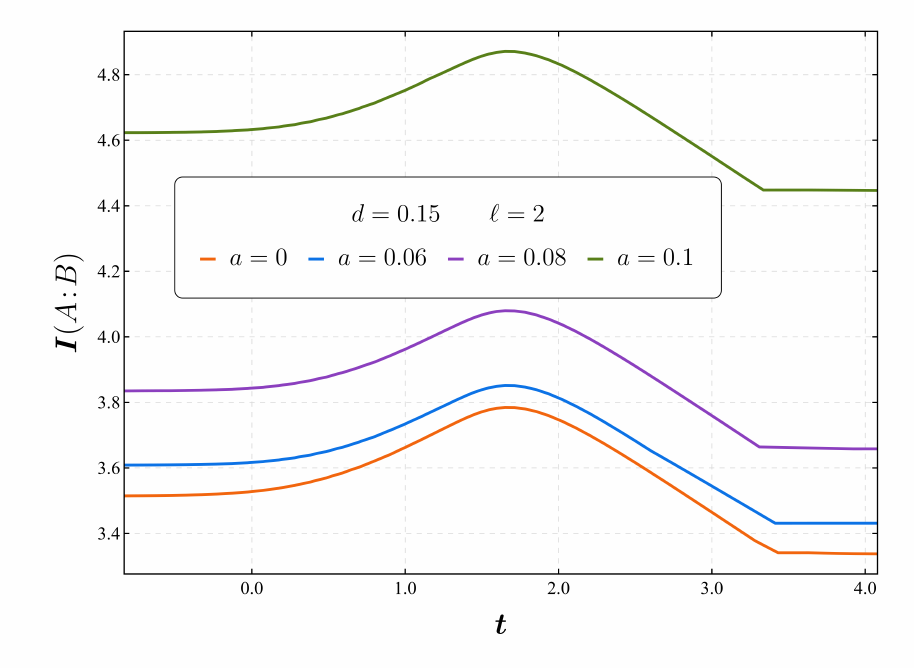}
			\par\smallskip
		\end{minipage}
		\hfill
		\begin{minipage}[t]{0.49\textwidth}
			\centering
			\includegraphics[width=\textwidth]
			{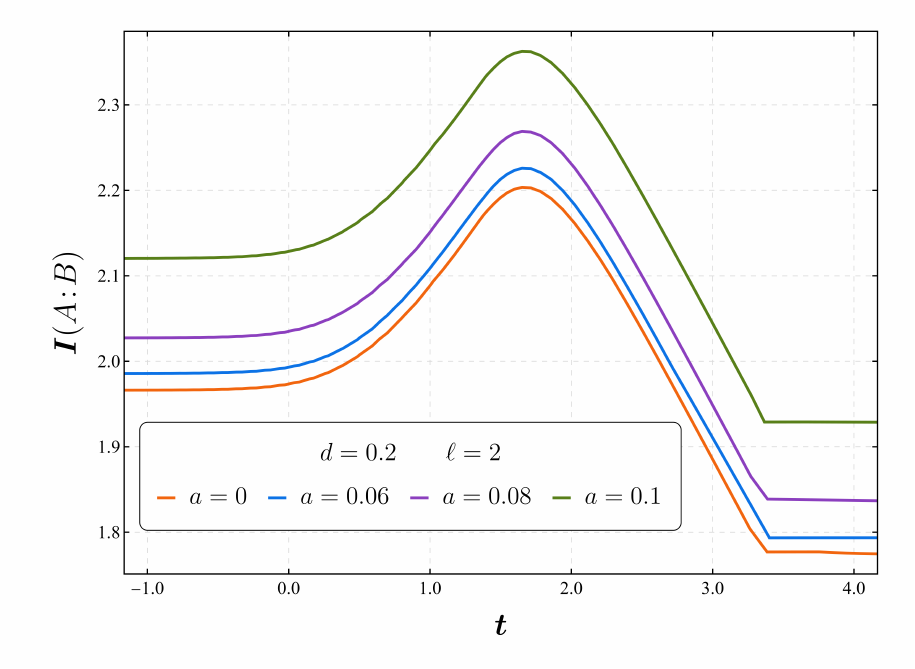}
			\par\smallskip
		\end{minipage}
		\caption{
			Evolution of the holographic mutual information $I(A:B)$ for several values of the parameters.
		}
		\label{fig:numerical-hmi-varying-a}
	\end{figure}

	In Figure~\ref{fig:numerical-hmi-varying-a} we show the evolution of the
	mutual information for several values of the parameters. We restrict
	ourselves to the range where the connected configuration is always
	dominant. At sufficiently early times, the mutual information approaches
	its vacuum value. It then increases, reaches a maximum, and decreases
	toward its final equilibrium value. For all curves shown, the equilibrium
	value is smaller than the vacuum value, since the thermal medium weakens
	the correlations. At fixed separation, the mutual information increases with the noncommutative scale throughout the evolution, with larger vacuum, maximum, and equilibrium values for larger \(a\) over the range shown. This trend is consistent with the static results of
	\cite{Fischler:2013gsa}. For the configurations shown in
	this figure, the mutual information remains
	positive at late times, so the transition to the disconnected
	configuration does not occur.

	\subsection{Entanglement wedge cross section}
\label{sec:ewcs}

Another holographic probe of correlations between two disjoint regions is the
entanglement wedge cross section (EWCS). For a connected entanglement
wedge in a static geometry, it is defined as the minimal
codimension-two cross section separating the portions of the wedge
associated with \(A\) and \(B\) \cite{Umemoto:2017aol,Nguyen:2017yqw}
\begin{equation}
	E_W(A:B)
	=
	\frac{\operatorname{Area}(\Sigma_{A\cup B})}{4G_N}.
	\label{eq:EWCS-prescription-intro}
\end{equation}
A notable property of this proposal is that the EWCS undergoes a
discontinuous phase transition as the separation between the two
subregions increases. As for the mutual information, the transition
originates from the competition between the connected and disconnected
entanglement-wedge configurations. When the separation is sufficiently
large and the disconnected configuration is favored, the entanglement
wedge is disconnected and \(E_W\) vanishes. This quantity has been
proposed as a holographic dual of the entanglement of purification, a
correlation measure for mixed states
\cite{Terhal:2002,Umemoto:2017aol,Nguyen:2017yqw}. Further geometric and thermal properties of the
EWCS have been investigated in a variety of
holographic backgrounds
\cite{BabaeiVelni:2019pkw,Jokela:2019ebz}.

\begin{figure}[t]
	\centering
	\includegraphics[width=0.86\textwidth]
	{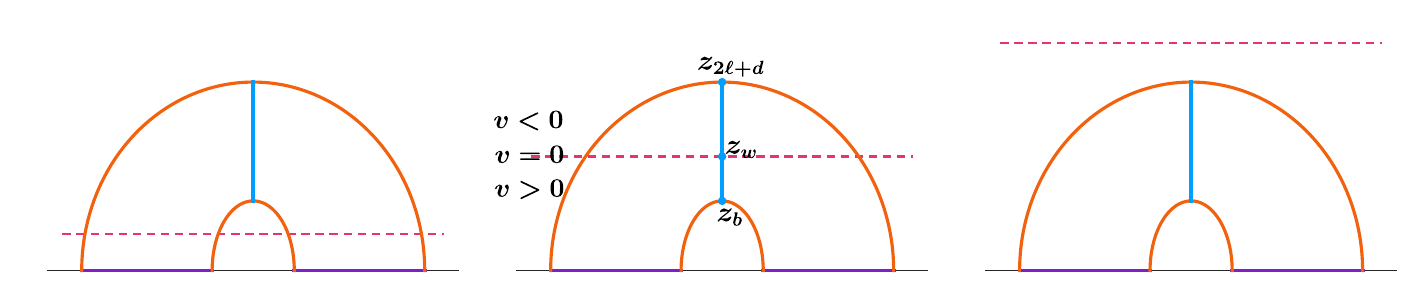}
	\caption{
		The three geometric configurations relevant for the EWCS evolution.
		The orange curves denote the HRT surfaces bounding the connected
		entanglement wedge, the blue curve denotes the EWCS, and the dashed
		line represents the infalling null shell. In the shell-crossing
		configuration the EWCS intersects the shell at \(z=z_w\).
	}
	\label{fig:ewcs-regimes}
\end{figure}

To find the evolution of EWCS we again consider the symmetric configuration depicted in
Figure~\ref{fig:hmi-configurations}. In the connected phase, the entanglement wedge is bounded by the HRT
surfaces associated with regions of widths \(d\) and \(2\ell+d\).
We denote the corresponding turning points by \(z_b\) and \(z_u\),
respectively. Reflection symmetry admits a symmetric cross-section candidate on the
plane \(x_2=0\), whose nontrivial profile lies in the \((v,z)\) plane.
During the evolution, three geometric configurations are relevant \cite{BabaeiVelni:2020wfl,BabaeiVelni:2023wqv} (see figure \ref{fig:ewcs-regimes}).
Before the null shell intersects the cross section, the connected EWCS
lies entirely in the vacuum region. At intermediate times, it crosses
the shell and contains vacuum and black-brane portions; the latter
may itself consist of two radial legs. Finally, if
the connected branch remains dominant, the cross section can enter
entirely into the final black-brane geometry.

To determine the connected cross section, we parameterize it by
\(v=v(z)\). Using the same normalization as in the previous sections, the corresponding area
functional is given by
\begin{equation}
	E_W(t)
	=
	\int_{z_b}^{z_u}
	\frac{dz}{z^3\sqrt{h(z)}}
	\sqrt{-f(v,z)v'(z)^2-2v'(z)},
	\label{eq:ewcs-functional}
\end{equation}
where a prime denotes differentiation with respect to \(z\). The
factor \(1/\sqrt{h(z)}\) follows from the noncommutative warp factor
together with the string-frame dilaton contribution and distinguishes
the present functional from its AdS counterpart. Extremizing Eq.~\eqref{eq:ewcs-functional} gives
\begin{equation}
	\frac{d}{dz}
	\left[
	\frac{1+f(v,z)v'}{
		z^3\sqrt{h(z)}
		\sqrt{-f(v,z)v'^2-2v'}}
	\right]
	=
	\frac{\partial_v f(v,z)\,v'^2}{
		2z^3\sqrt{h(z)}
		\sqrt{-f(v,z)v'^2-2v'}}.
	\label{eq:ewcs-eom}
\end{equation}
The profile is anchored on the two HRT surfaces through
\begin{equation}
	v(z_b)=v_b,
	\qquad
	v(z_u)=v_u .
	\label{eq:ewcs-boundary-conditions}
\end{equation}
Away from the null shell, \(\partial_v f=0\), and hence the momentum
conjugate to \(v\) is conserved separately in the two static regions.
We define the corresponding canonical momenta as
\begin{equation}
	\pi_\pm
	=
	-\frac{1+f_\pm v_\pm'}{
		z^3\sqrt{h(z)}
		\sqrt{-f_\pm v_\pm'^2-2v_\pm'}}.
	\label{eq:ewcs-momentum}
\end{equation}
Here the subscripts \(-\) and \(+\) refer to the vacuum region
\(v<0\) and the final black-brane region \(v>0\), respectively.
Solving Eq.~\eqref{eq:ewcs-momentum} for the profile gives
\begin{equation}
	v_\pm'(z)
	=
	-\frac{1}{f_\pm(z)}
	\left[
	1+
	\frac{\pi_\pm z^3\sqrt{h(z)}}
	{\sqrt{\mathcal V_\pm(z)}}
	\right],
	\qquad
	\mathcal V_\pm(z)
	=
	f_\pm(z)+\pi_\pm^2z^6h(z).
	\label{eq:ewcs-vprime}
\end{equation}
In the thin-shell limit, the extremal surface can be analyzed
in different regions of the Vaidya geometry, as discussed
separately below.

\textbf{Vacuum region (\(v<0\)).} Using Eq.~\eqref{eq:ewcs-vprime} and imposing the
boundary condition \(v(z_u)=v_u\), the vacuum profile becomes
\begin{equation}
	v_-(z)
	=
	v_u+z_u-z
	+
	\pi_-
	\int_z^{z_u}
	d\zeta\,
	\frac{\zeta^5}{
		\sqrt{\zeta^4+a^4+\pi_-^2\zeta^{10}}}.
	\label{eq:ewcs-vminus-profile}
\end{equation}
For a shell-crossing configuration, continuity of the profile at \(z=z_w\) requires \(v(z_w)=0\).

\textbf{Black-brane region (\(v>0\)).} In this case integrating Eq.~\eqref{eq:ewcs-vprime} and imposing
\(v(z_b)=v_b\), we find
\begin{equation}
	v_+(z)
	=
	v_b
	-
	\int_{z_b}^{z}
	\frac{d\zeta}{
		\sqrt{\mathcal V_+(\zeta)}
		\left[
		\sqrt{\mathcal V_+(\zeta)}
		-\pi_+\zeta^3\sqrt{h(\zeta)}
		\right]
	}.
	\label{eq:ewcs-vplus-profile}
\end{equation}
Using the matching condition at the shell the above equation gives
\begin{equation}
	v_b
	=
	\int_{z_b}^{z_w}
	\frac{d\zeta}{
		\sqrt{\mathcal V_+(\zeta)}
		\left[
		\sqrt{\mathcal V_+(\zeta)}
		-\pi_+\zeta^3\sqrt{h(\zeta)}
		\right]
	}.
	\label{eq:ewcs-black-continuity-regular}
\end{equation}
For a segment entirely outside the horizon, the same condition
may also be written in the separated form
\begin{equation}
	0
	=
	v_b+G(z_b)-G(z_w)
	-
	\pi_+
	\int_{z_b}^{z_w}
	d\zeta\,
	\frac{\zeta^5}{
		f(\zeta)
		\sqrt{
			f(\zeta)(\zeta^4+a^4)+\pi_+^2\zeta^{10}
	}},
	\label{eq:ewcs-black-continuity}
\end{equation}
where
\begin{equation}
	G(z)
	\equiv
	\int_{\delta}^{z}
	\frac{d\zeta}{f(\zeta)}.
	\label{eq:ewcs-G}
\end{equation}
Moreover, the endpoint conditions
for the vacuum and black-brane segments can also be expressed as
\begin{equation}
	v_b
	=
	-\int_{z_b}^{z_w}dz\,v_+'(z),
	\qquad
	v_u
	=
	\int_{z_w}^{z_u}dz\,v_-'(z).
	\label{eq:ewcs-endpoint-times}
\end{equation}
Since the inner endpoint belongs to the final static configuration, the corresponding boundary time satisfies
\begin{equation}
	t
	=
	v_b+G(z_b).
	\label{eq:ewcs-boundary-time-vb}
\end{equation}
Further, combining this relation with
Eq.~\eqref{eq:ewcs-black-continuity-regular} gives the
boundary time in terms of the shell-crossing
point,
\begin{equation}
	t
	=
	G(z_b)
	+
	\int_{z_b}^{z_w}
	\frac{dz}{
		\sqrt{\mathcal V_+(z)}
		\left[
		\sqrt{\mathcal V_+(z)}
		-\pi_+z^3\sqrt{h(z)}
		\right]
	}.
	\label{eq:ewcs-boundary-time}
\end{equation}

\textbf{Matching at the null shell.} Because \(f(v,z)\) jumps at \(v=0\), the conserved quantities on the
two sides of the shell are in general different. Integrating the
corresponding equation of motion through the infinitesimal shell gives
\begin{equation}
	\left.\frac{dz}{dv}\right|_+
	-
	\left.\frac{dz}{dv}\right|_-
	=
	\frac{g(z_w)}{2},
	\qquad
	g(z_w)=\left(\frac{z_w}{z_h}\right)^4 .
	\label{eq:ewcs-shell-jump}
\end{equation}
For the monotonic branch of Eq.~\eqref{eq:ewcs-vprime},
the above condition can be expressed as
\begin{equation}
	\sqrt{\mathcal V_+(z_w)}
	-\pi_+z_w^3\sqrt{h(z_w)}
	=
	\sqrt{\mathcal V_-(z_w)}
	-\pi_-z_w^3\sqrt{h(z_w)}.
	\label{eq:ewcs-matching}
\end{equation}
Solving this relation for $\pi_+$ gives
\begin{equation}
	\pi_+
	=
	\left(
	1-\frac{g(z_w)}{2}
	\right)\pi_-
	-
	\frac{g(z_w)}{
		2z_w^3\sqrt{h(z_w)}
	}
	\sqrt{
		1+\pi_-^2z_w^6h(z_w)
	}.
	\label{eq:ewcs-momentum-matching}
\end{equation}
Evaluating the contributions from the vacuum and black-brane
regions separately, the entanglement wedge cross section
takes the form
\begin{align}
E_W=
	\int_{z_b}^{z_w}
	\frac{dz}{
		z^3\sqrt{h(z)}
		\sqrt{\mathcal V_+(z)}
	}+	\int_{z_w}^{z_u}
	\frac{dz}{
		z^3\sqrt{h(z)}
		\sqrt{
			1+\pi_-^2z^6h(z)
	}}.
	\label{eq:ewcs-split}
\end{align}
When the cross section lies entirely in the vacuum region, the extremal surface has \(\pi_-=0\), and its area becomes
\begin{equation}
	E_W^{\mathrm{vac}}
	=
	\int_{z_b}^{z_u}
	\frac{dz}{
		z^3\sqrt{h(z)}
	},
	\label{eq:ewcs-vacuum}
\end{equation}

Before turning to the time evolution of the EWCS, it is useful
to examine several properties of \(\pi_-\) and \(\pi_+\)
that influence its behavior. In what follows, we mainly consider the large-region regime. We first examine the
approximation \(\pi_-\simeq0\), which considerably simplifies the
analytic treatment of \(E_W\). The null shell intersects the EWCS at \(z_w\) and the outer HRT surface at \(z_c\), which are generally distinct. Using \(z_c=z_u+v_u\) and Eq. \eqref{eq:ewcs-vminus-profile} we find
\begin{equation}
	z_w-z_c
	=
	\pi_-
	\int_{z_w}^{z_u}
	d\zeta\,
	\frac{\zeta^5}{
		\sqrt{\zeta^4+a^4+\pi_-^2\zeta^{10}}
	}.
	\label{eq:ewcs-zw-zc}
\end{equation}
In the large-region limit, where $z_w\simeq z_c\ll z_u$, the above result allows us to consider $\pi_-\simeq0$. Indeed, for fixed \(a\) and finite
\(z_w\), the integral in Eq.~\eqref{eq:ewcs-zw-zc} scales as \(z_u^4\) and hence
\(\pi_-=\mathcal O(z_u^{-4})\) in this limit. However, this approximation must also satisfy
the matching condition.
On the monotonic branch specified by
Eq.~\eqref{eq:ewcs-matching}, taking
$\pi_-\to0$ requires $g(z_w)\leq2$.
On the returning branch the sign of the square root is reversed, allowing $\pi_-\to0$ when
$g(z_w)>2$. In both cases the algebraic relation for
$\pi_+$ is Eq.~\eqref{eq:ewcs-momentum-matching},
provided the corresponding geometric root is chosen.
In large-region approximation, Eq.~\eqref{eq:ewcs-momentum-matching} gives
\begin{equation}
	\pi_+
	\simeq
	-\frac{g(z_w)}
	{2z_w^3\sqrt{h(z_w)}}
	=
	-\frac{\sqrt{z_w^4+a^4}}
	{2z_wz_h^4},
	\qquad
	(z_w\ll z_u).
	\label{eq:ewcs-large-region-momenta}
\end{equation}
For $g(z_w)>2$, the returning matching sign must be used; the
monotonic branch condition cannot be extrapolated
into that regime. In fact, in the $\pi_-\to0$ limit,
the matching condition with \(\dot z_-=-1\) gives \(\dot z_+=g(z_w)/2-1\).
If \(g(z_w)>2\), a curve that begins at the shell
and eventually reaches \(z_b<z_h<z_w\) must turn at
some \(z_r>z_w\); a globally monotonic \(z(v)\)
trajectory cannot meet both endpoint conditions. However, when \(z_w\simeq z_u\), we cannot assume
\(\pi_-\simeq0\). Indeed, in this situation
\(\pi_+\simeq0\) and \(\pi_-\neq0\). To see this, note that as
\(z_w\rightarrow z_u\), the outer HRT surface settles down to its final
static configuration as it lies entirely in the black-brane geometry,
and hence
\(t\simeq v_u+G(z_u)\). On the other hand, this expression must be consistent with
Eq.~\eqref{eq:ewcs-boundary-time} as \(z_w\simeq z_u\).
Since the final static turning point lies outside the horizon,
we obtain
\begin{equation}
	v_u
	=
	\pi_+
	\int_{z_b}^{z_u}
	d\zeta\,
	\frac{\zeta^5}{
		f(\zeta)
		\sqrt{
			f(\zeta)(\zeta^4+a^4)+\pi_+^2\zeta^{10}
	}}.
	\label{eq:ewcs-late-time-relation}
\end{equation}
As \(z_w\to z_u\), the shell-crossing condition implies \(v_u\to0\). Then the above equation gives \(\pi_+\to0\).
In this situation, solving Eq.~\eqref{eq:ewcs-momentum-matching} for \(\pi_+=0\) gives
\begin{equation}
	\pi_-^2
	\simeq
	\frac{g(z_u)^2}
	{4f(z_u)z_u^6h(z_u)}
	=
	\frac{z_u^4+a^4}
	{4z_u^2z_h^4\left(z_h^4-z_u^4\right)},
	\qquad
	(z_w\simeq z_u).
	\label{eq:ewcs-late-momenta}
\end{equation}
As we will argue, this result allows us to explore the late-time
behavior of \(E_W\) when the connected cross section lies entirely in
the final black-brane region.
\begin{figure}[t]
	\begin{center}
		\includegraphics[scale=0.36]
		{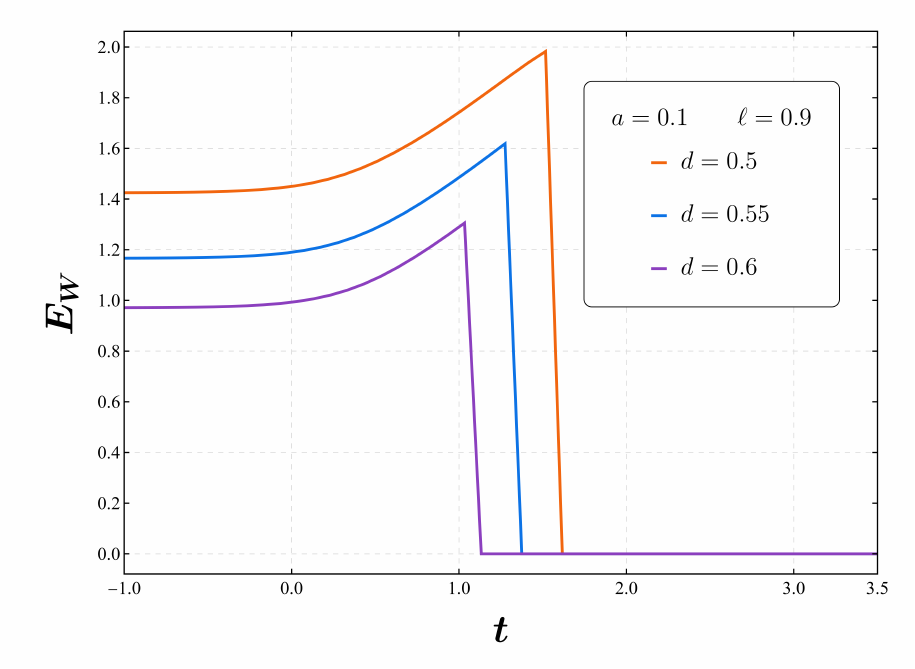}
		\hspace*{-0.4cm}
		\includegraphics[scale=0.36]
		{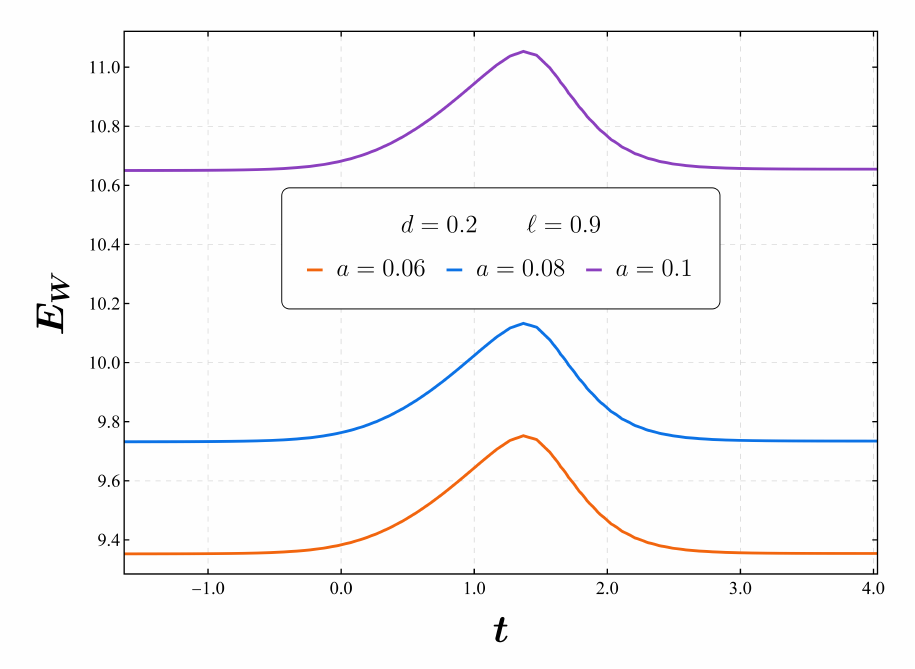}
		\hspace*{-0.4cm}
		\includegraphics[scale=0.36]
		{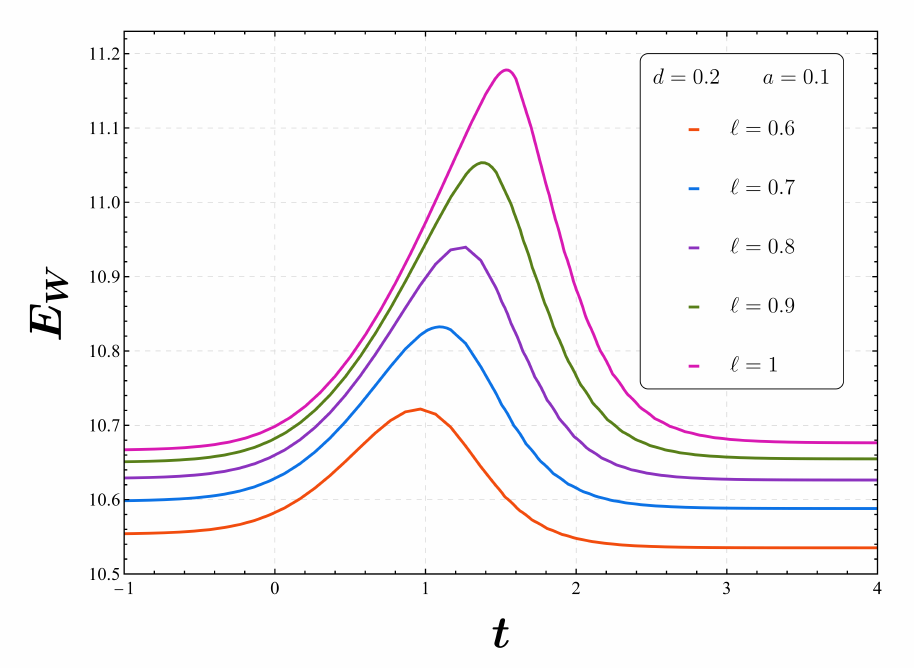}
	\end{center}
	
	\caption{
		Time evolution of the entanglement wedge cross section for several
		values of the parameters.
	}
	\label{fig:numerical-ewcs-comparison}
\end{figure}

The numerical results for the time evolution of the EWCS are
presented in Fig.~\ref{fig:numerical-ewcs-comparison}. The left
panel shows that increasing the strip separation suppresses
$E_W$ and can induce a discontinuous transition of the
connected configuration to a vanishing EWCS. The middle panel
illustrates its dependence on the noncommutativity scale,
revealing a transient enhancement as the deformation increases.
The right panel compares different strip widths and shows that
the EWCS approaches a finite equilibrium value whenever the
connected configuration remains dominant.

\begin{figure}[t]
	\centering
	\begin{minipage}[t]{0.49\textwidth}
		\centering
		\includegraphics[width=\linewidth]
		{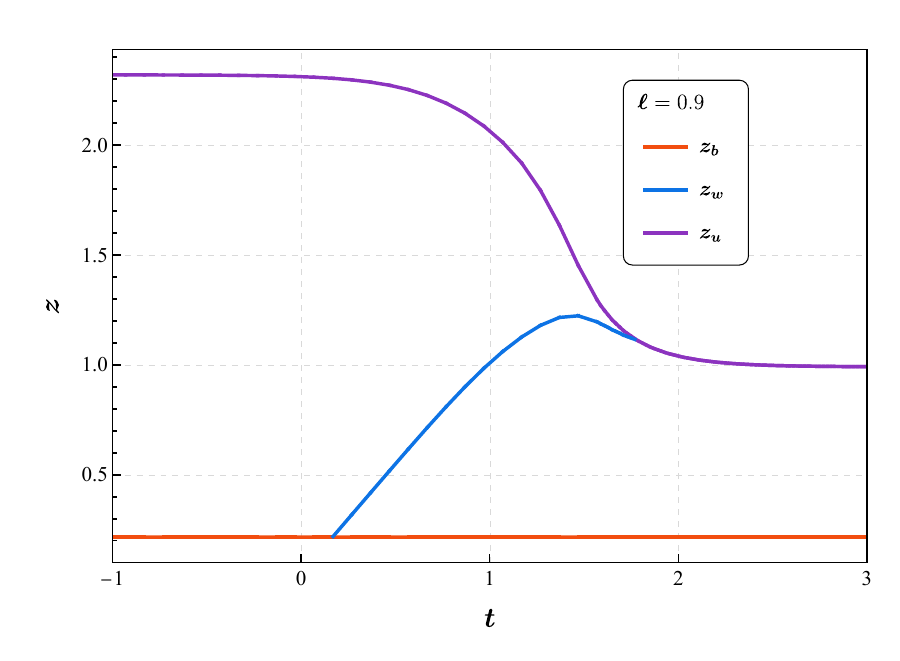}
	\end{minipage}
	\hfill
	\begin{minipage}[t]{0.47\textwidth}
		\centering
		\includegraphics[width=\linewidth]
		{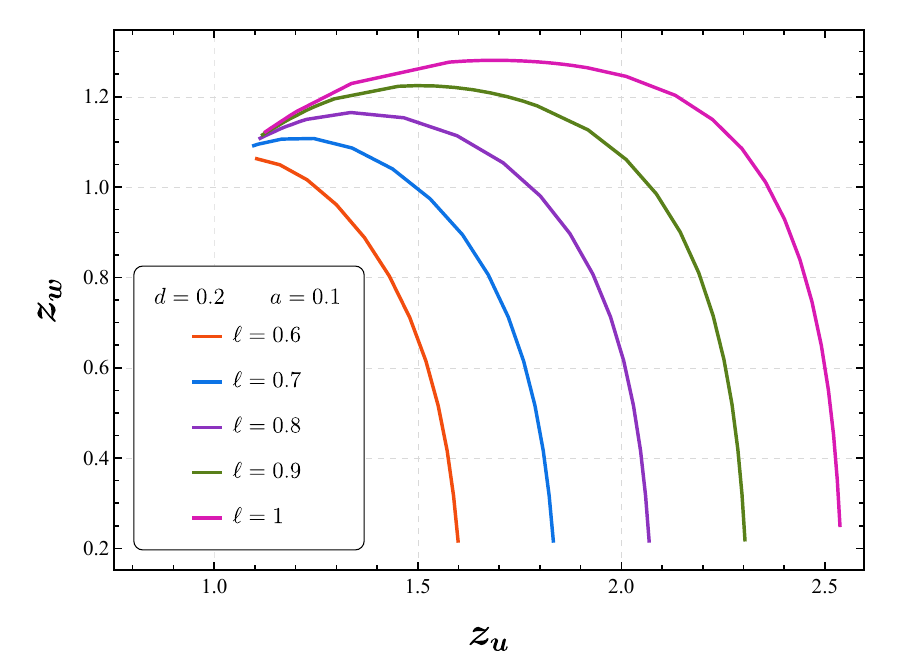}
	\end{minipage}
	
	\caption{
		\textit{Left}: Time evolution of $z_b$, $z_w$, and $z_u$.
		\textit{Right}: Parametric curves $(z_u(t), z_w(t))$ for several strip widths.
	}
	\label{fig:ewcs-radial-positions}
\end{figure}

To further elucidate the time evolution of the EWCS, we
examine the radial positions \(z_b\), \(z_w\), and \(z_u\).
The left panel of Fig.~\ref{fig:ewcs-radial-positions}
shows their evolution with boundary time. The shell-crossing
position \(z_w\) is displayed only when the EWCS intersects
the null shell. The inner turning point \(z_b\) remains
nearly constant, whereas the outer turning point \(z_u\)
decreases toward its final equilibrium value. During the
shell-crossing regime, \(z_w\) initially increases and then
approaches \(z_u\), while remaining within the interval
\(z_b<z_w<z_u\). As \(z_w\) approaches \(z_u\), the vacuum
contribution to the EWCS decreases and eventually vanishes.
This behavior indicates that, for the configurations
considered, the evolution of the EWCS is governed primarily
by the outer HRT surface. The right panel of Fig.~\ref{fig:ewcs-radial-positions}
displays the corresponding parametric relation between
\(z_w\) and \(z_u\), with boundary time serving as an
implicit parameter. The resulting trajectories are
non-monotonic, and increasing the strip width extends them
toward larger radial positions.

The closed-form expression for $E_W$ in Eq.~\eqref{eq:ewcs-split}
allows its scaling behavior to be extracted in different stages
of the thermalization process. In the large-\(\ell\) limit, as
in the case of entanglement entropy, the time evolution of the
EWCS exhibits three distinct regimes: early-time growth,
intermediate linear growth, and late-time saturation.
These regimes are examined in turn below.

\textbf{Early-time growth.}
At sufficiently early times, the infalling shell does not
intersect the EWCS, which therefore retains its vacuum value.
The initial evolution begins once the shell reaches the cross
section. Following the approach of \cite{BabaeiVelni:2023wqv},
the shell-crossing position is parametrized as
\[
    z_w=z_b+\eta,
    \qquad 0<\eta\ll z_b,
\]
with the large-region hierarchy
\(z_b\ll z_h\ll z_u\). In this regime, the preceding analysis
yields \(\pi_-\simeq0\) and
\(\pi_+=\mathcal{O}(z_h^{-4})\), at fixed \(a\) and \(z_w\).
Expanding Eq.~\eqref{eq:ewcs-split} then gives the leading
thermal correction
\begin{equation}
    \Delta E_W(t)
    = E_W(t)-E_W^{\mathrm{vac}}
    \simeq
    \frac{1}{2z_h^4}
    \int_{z_b}^{z_w} dz\,
    \frac{\sqrt{z^4+a^4}}{z}.
    \label{eq:ewcs-early-integral}
\end{equation}
This expression applies near the onset of shell crossing,
where \(z_w-z_b\ll z_b\). Moreover, expanding Eq.~\eqref{eq:ewcs-boundary-time} at
early times gives
\begin{equation}
    t=z_w-\delta+
    \mathcal O\left(\frac{z_w^5}{z_h^4}\right).
    \label{eq:ewcs-early-time-relation}
\end{equation}
The inner turning point \(z_b\) is determined by the strip
separation \(d\). To leading order in \(z_b/z_h\), the
corresponding vacuum width relation reads
\begin{equation}
    d=2\int_{\delta}^{z_b} dz\,
    \frac{z\sqrt{z^4+a^4}}
    {\sqrt{z_b^6-z^6}}.
    \label{eq:ewcs-early-width}
\end{equation}
In the limit \(\delta\ll a\ll z_b\), the
corresponding expansion takes the form
\begin{equation}
    d\simeq
    \mathcal B z_b
    +\frac{a^4}{z_b^3}
    \left[
        \log\left(\frac{z_b}{a}\right)
        +\frac14+\frac56\log 2
    \right]
    +\mathcal O\left(\frac{a^8}{z_b^7}\right),
\end{equation}
with
\(
    \mathcal B=\frac13 B\left(\frac23,\frac12\right)
\) and $B(x,y)$ is the Euler beta function. Considering the limit
$z_b\lesssim t\ll z_h$ and expanding the integrand of
Eq.~\eqref{eq:ewcs-early-integral} in powers of $a^4/z^4$, at leading order, we obtain
\begin{equation}
	\Delta E_W(t,d)\simeq
	\frac{\pi A_\Sigma}{6}\mathcal E
	\Bigg[
	t^2
	-\frac{a^4}{2 t^2}
	+\cdots
	\Bigg].
	\label{eq:ewcs-early-general-energy-density}
\end{equation}
Therefore at early times EWCS grows quadratically and the rate of growth is a fixed constant proportional to the energy density. Note that the correction at order $a^4$ receives contributions from both the radial integrand
and the noncommutative shift of the turning point at fixed separation. In the commutative limit, the above result agrees with the early time scaling of EWCS in relativistic setup \cite{BabaeiVelni:2020wfl}.\footnote{Note that in the limit \(z_b\ll a\), with the UV cutoff satisfying \(\delta\ll z_b^3/a^2\), the leading contribution takes the logarithmic form \(\Delta E_W(t,d)\simeq \frac{\pi A_\Sigma}{3}\mathcal E\,a^2\log\left(\frac{t d}{\mathcal C a^2}\right)\). However, this expression is obtained outside the physical regime \(d>\ell_c\). Indeed, combining \(d\simeq\mathcal C a^2/z_b\) with \(\ell_c\simeq a^2/(2\delta)\) gives \(d/\ell_c\simeq 2\mathcal C\delta/z_b\ll1\) for \(\delta\ll z_b\). Thus, the logarithmic expression does not correspond to a physically resolvable regime of the finite-cutoff theory.
}

\textbf{Intermediate time.}
We now consider the intermediate-time regime
\(d\ll z_h\ll t\ll\ell\), where
\(z_b\ll z_w\ll z_u\).
For the returning branch, the black-brane segment has
a radial turning point \(z_r\), where
\(\mathcal V_+(z_r)=0\) and
\(z_h<z_w<z_r<z_m\).
The EWCS integral and the regular boundary-time relation
obtained above are modified by the inclusion of the
additional returning segment
\begin{equation}
    \begin{aligned}
        E_W^+
        ={}&
        \int_{z_b}^{z_w}
        \frac{dz}{z^3\sqrt{h(z)}\sqrt{\mathcal V_+(z)}}
        +2\int_{z_w}^{z_r}
        \frac{dz}{z^3\sqrt{h(z)}\sqrt{\mathcal V_+(z)}},
        \\[5pt]
        t
        ={}& G(z_b)
        +\int_{z_b}^{z_w}
        \frac{dz}{
        \sqrt{\mathcal V_+(z)}
        \left[\sqrt{\mathcal V_+(z)}
        -\pi_+z^3\sqrt{h(z)}\right]}
        +2\int_{z_w}^{z_r}
        \frac{\pi_+z^3\sqrt{h(z)}\,dz}
        {f(z)\sqrt{\mathcal V_+(z)}}.
    \end{aligned}
    \label{eq:ewcs-intermediate-integrals}
\end{equation}
The corresponding expression for a single shell crossing
is given in Eq.~\eqref{eq:ewcs-boundary-time};
the second integrals above account for the additional
returning segment.
Along this segment, the local momentum conjugate to \(v(z)\)
changes sign, from \(\pi_+\) to \(-\pi_+\), while
\(\pi_+<0\) continues to denote the constant Hamiltonian
in the \(z(v)\) parametrization.
The matching condition for the returning branch reads
\[
    \sqrt{\mathcal V_+(z_w)}+\pi_+q(z_w)
    =
    -\left[
    \sqrt{\mathcal V_-(z_w)}-\pi_-q(z_w)
    \right],
    \qquad
    q(z)=z^3\sqrt{h(z)}.
\]
In the large-region
limit \(\pi_-\to0\), it implies
\[
    \pi_+q(z_w)
    =-\frac12\left(\frac{z_w}{z_h}\right)^4,
    \qquad
    \sqrt{\mathcal V_+(z_w)}
    =\frac12\left(\frac{z_w}{z_h}\right)^4-1.
\]
Consequently, the returning shell-crossing configuration
requires \(z_w>2^{1/4}z_h\). Following the near-critical analysis of
\cite{BabaeiVelni:2020wfl,BabaeiVelni:2023wqv},
we consider a critical configuration in which
$\mathcal V_+$ develops a double zero at $z=z_m$,
\[
\mathcal V_+(z_m,\pi_\star)=0,
\qquad
\left.\partial_z\mathcal V_+\right|_{z_m,\pi_\star}=0,
\]
where $\pi_\star=\pi_+|_{z_w=z_w^\star}$.
The returning solutions approach this configuration
from $\pi_+>\pi_\star$, with $z_r\to z_m^-$.
Expanding $\mathcal V_+$ around the critical point gives
\begin{equation}
	\mathcal V_+(z,\pi_+)
	\simeq
	c_1(z-z_m)^2
	+c_2(\pi_\star-\pi_+),
	\label{eq:ewcs-critical-expansion}
\end{equation}
where
\[
c_1:=
\frac12
\left.
\partial_z^2\mathcal V_+
\right|_{z_m,\pi_\star},
\qquad
c_2:=
-\left.
\partial_{\pi_+}\mathcal V_+
\right|_{z_m,\pi_\star}
=-2\pi_\star z_m^6h(z_m)>0,
\]
and
\(
\pi_\star:=
-\frac{1}{z_m^3}
\sqrt{-\frac{f(z_m)}{h(z_m)}}<0
\). The critical radius $z_m$ coincides with that obtained
from the HRT analysis in Eq.~\eqref{eq:critical-zm}.
At this radius, the double zero of $\mathcal V_+$
is a nondegenerate minimum, so $c_1>0$.
Consequently, as $\pi_+\to\pi_\star^+$,
the returning turning point approaches $z_m$ as
\[
z_r=z_m-
\sqrt{\frac{c_2(\pi_+-\pi_\star)}{c_1}}
+\cdots.
\]
Combining the shell-matching condition with
Eq.~\eqref{eq:critical-zm}, the critical configuration
is found to satisfy
\[
    2^{1/4}z_h<z_w^\star<z_m.
\]
Thus, the single-crossing segment terminates before
reaching the critical radius, whereas the returning
segment extends toward the double zero, giving rise
to the logarithmic divergence and the resulting
linear growth of the EWCS. Using Eq.~\eqref{eq:ewcs-critical-expansion},
the leading contribution to the $E_W$ is
\begin{equation}
	E_W
	\simeq
	\frac{2}{z_m^3\sqrt{h(z_m)}}
	\int_{z_w}^{z_r}
	\frac{dz}{\sqrt{\mathcal V_+(z)}}
	+\mathcal O(1).
	\label{eq:ewcs-critical-ew}
\end{equation}
The vacuum contribution and the non-returning
black-brane segment remain finite in the critical
limit, so the logarithmic divergence arises
entirely from the returning segment. Similarly, the leading contribution to the
boundary time follows from
Eq.~\eqref{eq:ewcs-intermediate-integrals},
\begin{equation}
	t
	\simeq
	\frac{2\pi_\star z_m^3\sqrt{h(z_m)}}{f(z_m)}
	\int_{z_w}^{z_r}
	\frac{dz}{\sqrt{\mathcal V_+(z)}}
	+\mathcal O(1).
	\label{eq:ewcs-critical-time}
\end{equation}
The EWCS and boundary time share the same leading
logarithmic divergence. Taking the ratio of their
coefficients gives the reduced growth rate
$-\pi_\star=v_E^{\mathrm{NC}}/z_h^3$ and hence the intermediate scaling of the EWCS becomes
\begin{equation}
	E_W
	\simeq
	\frac{A_\Sigma}{2}s_{\mathrm{th}}
	v_E^{\mathrm{NC}}t+\cdots,
	\label{eq:ewcs-linear-growth}
\end{equation}
where $s_{\mathrm{th}}$ is the thermal entropy density
defined in Eq.~\eqref{eq:setup-thermodynamics},
$A_\Sigma=2L^2$, and
\begin{equation}
	v_E^{\mathrm{NC}}
	=
	\frac{z_h^3}{z_m^3}
	\sqrt{
		-\frac{f(z_m)}{h(z_m)}
	}.
	\label{eq:ewcs-velocity}
\end{equation}
The resulting EWCS growth velocity coincides with
the entanglement velocity obtained from the HRT
analysis in Eq.~\eqref{eq:linear-entropy-growth}.
The linear growth is governed by the returning
black-brane segment, which probes the critical
region near $z_m$.

\textbf{Saturation.}
At late times, provided the connected entanglement wedge
remains dominant, the EWCS approaches its equilibrium value
in the final black-brane geometry. As the shell-crossing
position approaches its static limit, \(z_w\to z_u\), the
vacuum contribution vanishes, \(E_W^-\to0\). In this limit,
the static configuration satisfies \(v'(z)=-1/f(z)\),
implying that the conserved momentum vanishes,
\(\pi_+=0\), and hence \(\mathcal V_+(z)=f(z)\).
Restoring the transverse and gravitational normalization,
Eq.~\eqref{eq:ewcs-split} yields
\begin{equation}
    E_W
    =
    \frac{A_\Sigma}{2}s_{\mathrm{th}}z_h^3
    \int_{z_b}^{z_u}
    \frac{dz}{z^3\sqrt{h(z)f(z)}},
    \label{eq:ewcs-late-equilibrium}
\end{equation}
which coincides witht the results first reported in \cite{Chowdhury:2021hlz}.
In particular, in the low-temperature, small $a$ limit, the EWCS becomes 
\begin{equation}
			E_W(\ell,d)
				\simeq
				\frac{A_\Sigma \mathcal B^2}{4}s_{\mathrm{th}}z_h^3
				\Bigg[
				\frac1{d^2}
				-\frac1{(2\ell+d)^2}
				+\frac{a^4\mathcal B^4}{6}
				\left(
				\frac1{d^6}
				-\frac1{(2\ell+d)^6}
				\right)+\cdots
				\Bigg].	
\end{equation}			
If the disconnected HRT configuration instead becomes
dominant, the entanglement wedge disconnects and the
physical EWCS vanishes. The transition to this configuration
is generically discontinuous
\cite{BabaeiVelni:2020wfl}.

\section{Conclusions and discussions}
\label{sec:conclusion}

We have studied the time evolution of holographic information measures following a homogeneous global quench in four-dimensional $\mathcal{N}=4$ noncommutative Yang--Mills theory. The quench is modeled by an infalling null shell that changes the energy density at fixed noncommutative scale. We consider strip regions whose finite direction lies along the noncommutative coordinate $x_2$, combining analytic results in the thin-shell and large-region limits with numerical calculations for a smooth shell. The corresponding noncommutative Vaidya geometry is supported by an external null source and satisfies the sourced type IIB equations, as demonstrated in Appendix~\ref{app:nc-vaidya-consistency}.

Noncommutativity affects the evolution through the non-AdS near-boundary geometry and its associated UV/IR mixing. Already in the static problem, the U-shaped extremal surfaces exhibit a local minimum width $\ell_0$, separating two branches, as well as a cutoff-dependent scale $\ell_c$ associated with UV/IR mixing \cite{Fischler:2013gsa}. For $\delta\ll a$, the latter behaves as $\ell_c\simeq \frac{a^2}{2\delta}$.
We therefore restrict our dynamical analysis to the resolvable regime above $\ell_c$. In the parameter range considered, the selected vacuum and equilibrium surfaces lie on the deeper U-shaped branch.

At early times, the leading near-boundary contribution to the entanglement entropy is logarithmic, with a coefficient proportional to the injected energy density and to $a^2$. At this order, the contribution is independent of the strip width. This result holds within the hierarchy specified in Sec.~\ref{subsec:early-growth}; width-dependent terms enter beyond the leading near-boundary expansion. Moreover, the commutative limit of this expansion is nonuniform, and the commutative result must instead be recovered from the full expressions.

For sufficiently wide strips, the evolution is governed by the critical geometry behind the final horizon, as in the standard holographic quench scenario \cite{Liu:2013iza}. In particular, an intermediate linear-growth regime emerges from the double zero of the effective potential at $z=z_m$. At fixed $z_h$, both $z_m$ and the critical shell-crossing radius $z_c^\ast$ move toward the horizon as $a$ increases over the range shown in Figure~\ref{fig:critical-scales-a}, while the ordering $z_h<z_c^\ast<z_m$ is maintained. The corresponding entanglement velocity $v_E^{\mathrm{NC}}$ approaches the AdS$_5$ result in the commutative limit and increases with $a$ over the range studied, in line with the previously observed effects of noncommutativity on entanglement propagation \cite{Fischler:2018kwt}.

The late-time analysis reveals a distinction between the entanglement velocity and the butterfly velocity. As shown in Eq.~\eqref{eq:saturation-butterfly-time}, the leading large-strip behavior of the continuous-completion time $t_s^{(\mathrm{cont})}$ is controlled by $v_{B,x_2}$, whereas the intermediate linear growth is governed by $v_E^{\mathrm{NC}}$. The continuous-completion scale should not, in general, be identified with the physical saturation time. Different HRT branches may compete in the noncommutative theory, and the physical entropy is determined by the globally minimal branch.

The numerical results are consistent with this picture. At fixed cutoff, increasing the noncommutative scale raises the equilibrium value of the entropy difference and modestly increases the slope in the intermediate growth regime; the change in the equilibrium value is more pronounced than that in the growth rate. Increasing the strip width primarily extends the growth regime and delays saturation, while leaving the leading growth rate nearly width independent. For wider strips, the change in slope near the plateau is consistent with a change in the dominant HRT branch. The numerical profiles also indicate that, at fixed width, varying the noncommutative scale affects the turning point less strongly than varying the strip width.

\begin{figure}[t]
    \centering
    \includegraphics[width=0.55\textwidth]
    {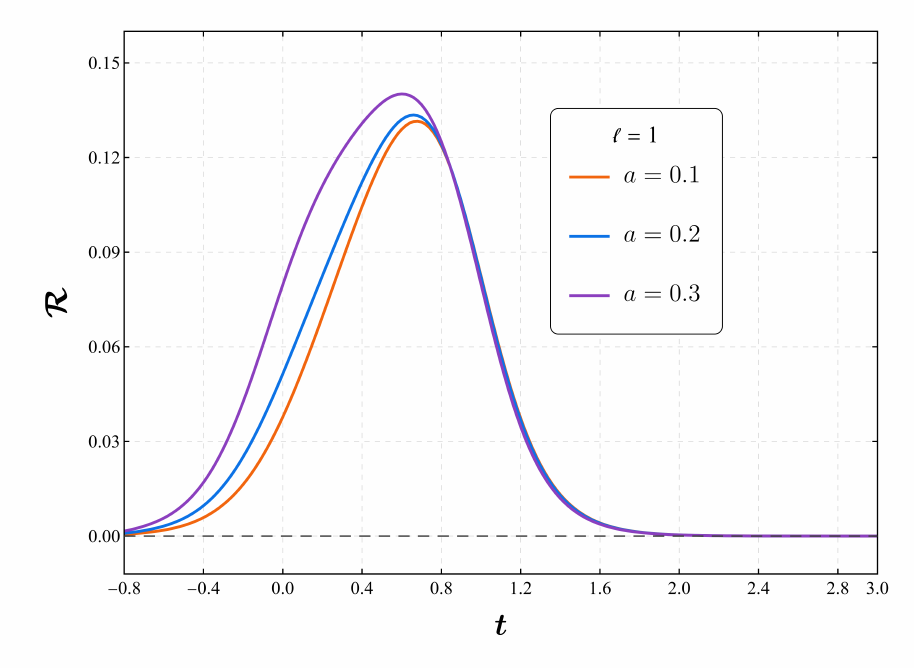}
    \caption{
       $\mathcal R(t)$
        for different values of the noncommutative scale.
    }
    \label{fig:numerical-entropy-rate}
\end{figure}

To further characterize the time dependence, we consider the instantaneous entropy growth rate, following the holographic quench analysis of \cite{Liu:2013qca}. With the normalization of Eq.~\eqref{eq:setup-entropy-ten-dimensional}, we define
\begin{equation}
    \mathcal R(t;\ell)
    =\frac{1}{s_{\rm th} A_{\Sigma}}\frac{dS(\ell,t)}{dt},
    \label{eq:linear-entropy-rate}
\end{equation}
where in the intermediate linear-growth regime reduces to the entanglement velocity. Figure~\ref{fig:numerical-entropy-rate} shows the instantaneous rate for a strip of width $\ell=1$. For all three values of the noncommutative scale, the rate rises to a single maximum and subsequently decreases toward zero as the entropy approaches equilibrium. Increasing $a$ shifts the maximum to earlier times and modestly increases its height.

We have also investigated the holographic mutual information between two disjoint strips. The leading width-independent logarithmic contribution at early times cancels in the connected combination. Consequently, determining the leading time-dependent correction requires the subleading, width-dependent terms. For sufficiently large regions whose linear-growth regimes overlap, the mutual-information growth rates are inherited from the physical single-strip entropies and are governed by the same entanglement velocity $v_E^{\mathrm{NC}}$. This is consistent with previous holographic studies of mutual information following a quench \cite{Allais:2011ys,Alishahiha:2014jxa,Tanhayi:2015cax}. The numerical HMI curves begin at their vacuum values, rise to a maximum, and then decrease toward equilibrium. Increasing the noncommutative scale raises the mutual information throughout the evolution which is consistent with the static NCYM result of \cite{Fischler:2013gsa}.

We next considered the entanglement wedge cross section. Its vacuum and equilibrium expressions reproduce the known static NCYM results \cite{Chowdhury:2021hlz}. On the near-critical returning branch, the EWCS extremal candidate probes the same critical radius $z_m$ as the HRT surface and exhibits linear growth governed by the same entanglement velocity $v_E^{\mathrm{NC}}$. The numerical results exhibit two possible late-time behaviors. If the connected entanglement wedge becomes subdominant, the physical EWCS vanishes; increasing the separation then leads to earlier disentangling. If the connected phase remains dominant, the EWCS develops a transient maximum and subsequently approaches a finite equilibrium value. Again, we observed that increasing the noncommutative scale raises the vacuum, maximum, and equilibrium values.

Although our analysis has focused on the entanglement entropy, mutual information, and EWCS, the latter is conjectured to be dual to several information-theoretic measures of mixed-state correlations. It is therefore interesting to briefly examine what our results imply for these quantities. In particular, the time evolution of the EWCS, together with the entanglement entropies calculated above, allows us to infer the behavior of reflected entropy, the Markov gap, and odd entanglement entropy. These quantities provide complementary information-theoretic perspectives on the results obtained in this work.

The reflected entropy \(S_R(A:B)\) is defined through the canonical purification of the reduced density matrix \(\rho_{AB}\) \cite{Dutta:2019gen}. At leading order in holography, it is related to the EWCS by
\begin{equation}
    S_R(A:B;t)=2E_W(A:B;t).
    \label{eq:reflected-entropy-EWCS-intro}
\end{equation}
Reflected entropy has also been studied in anisotropic holographic theories \cite{Vasli:2022anisotropic}, in covariant time-dependent settings \cite{Afrasiar:2022}, and following global quenches in two-dimensional conformal field theories \cite{Moosa:2020vcs}.

The Markov gap \(\mathcal M_g(A:B;t)\) is defined as the difference between reflected entropy and mutual information. It is nonnegative and is associated with quantum Markov recovery \cite{Hayden:2021gno}. Using Eq.~\eqref{eq:reflected-entropy-EWCS-intro}, we obtain
\begin{equation}
    \begin{aligned}
        \mathcal M_g(A:B;t)
        \equiv S_R(A:B;t)-I(A:B;t)
        =2E_W(A:B;t)-I(A:B;t)
        \geq 0.
    \end{aligned}
    \label{eq:markov-gap-EWCS}
\end{equation}
Thus, the Markov gap provides a measure of the difference between the correlations captured by reflected entropy and those quantified by mutual information.

We also consider the odd entanglement entropy \(S_O(A:B)\), defined through the odd-replica continuation of the partial transpose of \(\rho_{AB}\) \cite{Tamaoka:2018ned}. Adopting the proposed leading-order holographic relation, we write
\begin{equation}
    S_O(A:B;t)
    =
    S(A\cup B;t)+E_W(A:B;t).
    \label{eq:odd-entropy-EWCS}
\end{equation}
Here, \(S(A\cup B;t)\) is determined by the competing connected and disconnected HRT configurations discussed in Eq.~\eqref{eq:hmi-union-entropy}. In evaluating \(S_O\), we subtract the UV-divergent contribution from \(S(A\cup B;t)\), while leaving the EWCS contribution unchanged.

\begin{figure}[t]
    \centering
    \begin{minipage}[t]{0.32\textwidth}
        \centering
        \includegraphics[width=\linewidth]
        {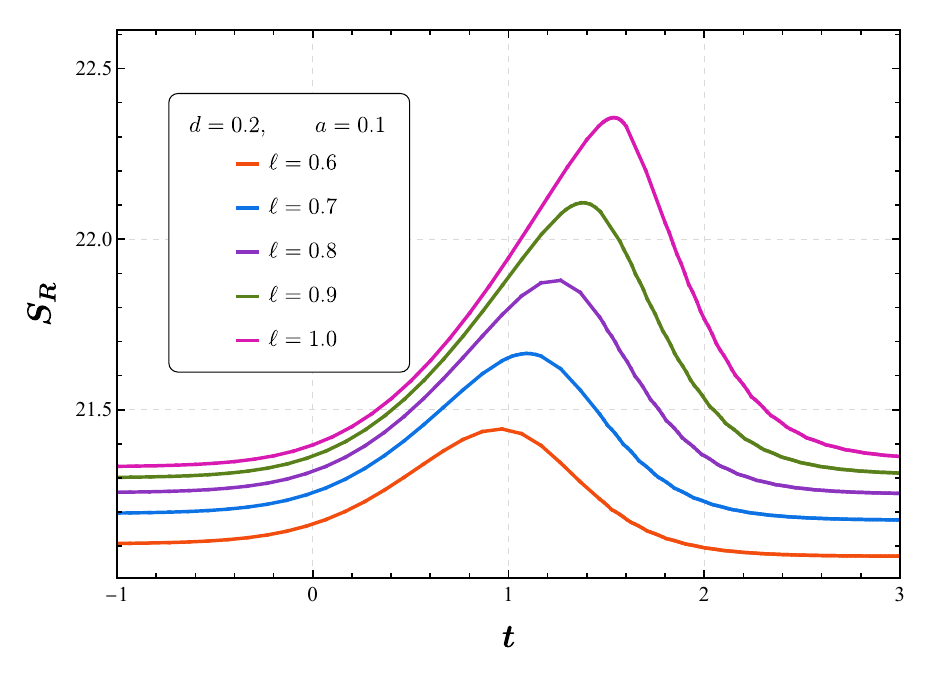}
    \end{minipage}
    \hfill
    \begin{minipage}[t]{0.32\textwidth}
        \centering
        \includegraphics[width=\linewidth]
        {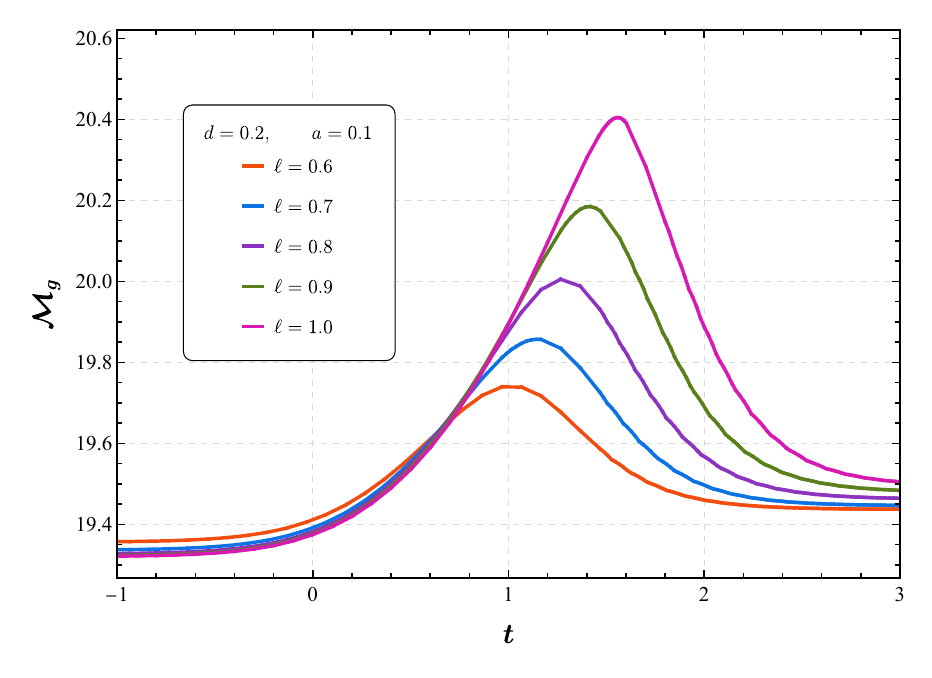}
    \end{minipage}
    \hfill
    \begin{minipage}[t]{0.32\textwidth}
        \centering
        \includegraphics[width=\linewidth]
        {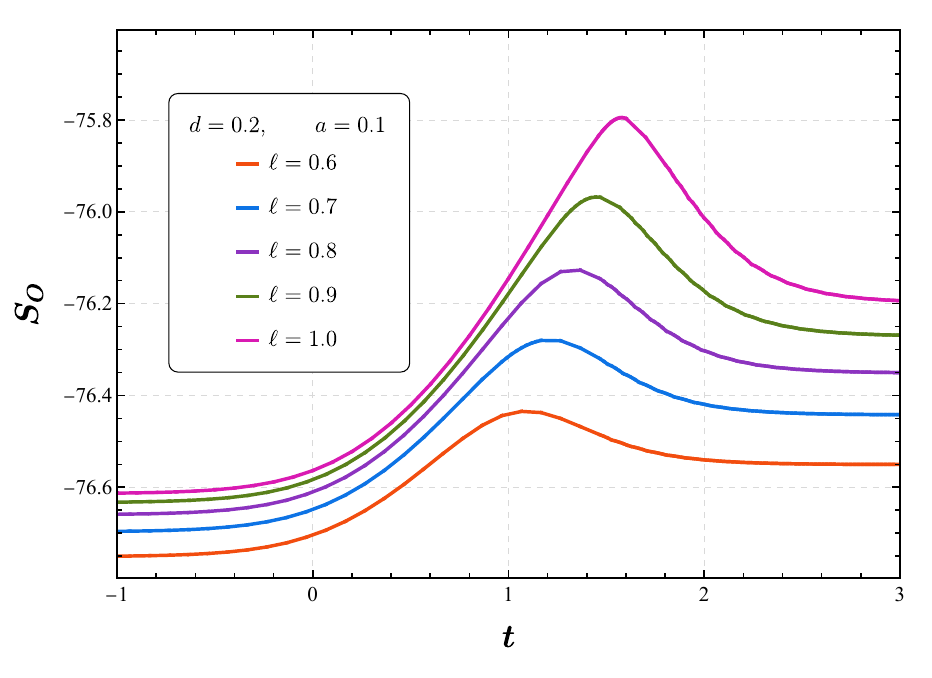}
    \end{minipage}
    \caption{
        Time evolution of three boundary-theory measures of
        mixed-state correlations for different strip widths
        at fixed \(d=0.2\) and \(a=0.1\).
        \textit{Left}: reflected entropy \(S_R\).
        \textit{Middle}: Markov gap \(\mathcal M_g\).
        \textit{Right}: UV-subtracted odd entanglement entropy \(S_O\).
    }
    \label{fig:numerical-reflected-markov-odd}
\end{figure}

Figure~\ref{fig:numerical-reflected-markov-odd} illustrates the time evolution of these three quantities for several parameters.
As shown in the left panel the reflected entropy directly inherits the time dependence of the EWCS. The middle panel shows that the Markov gap develops a transient maximum. As the strip width increases, this maximum becomes larger and shifts to later times, reflecting the different time dependences of the EWCS and mutual information. The right panel shows that the odd entanglement entropy has a more involved evolution, since it receives contributions from both the EWCS and the entropy of the union \(A\cup B\).
An interesting feature emerges when the disconnected HRT configuration becomes dominant. In this regime, the entanglement wedge disconnects, and both \(E_W\) and \(I(A:B)\) vanish at leading order. Consequently, the reflected entropy and Markov gap also vanish, whereas the odd entanglement entropy reduces to \(S(A\cup B;t)\). These observations illustrate that, although all three quantities are related to the EWCS, they capture different aspects of mixed-state correlations and provide complementary perspectives on the nonequilibrium dynamics.

Several extensions of this work would be worthwhile. First, it would be interesting to investigate the dynamics of small entangling regions, along the lines of \cite{Kundu:2016cgh}, and to determine whether the corresponding behavior can be reproduced directly through field-theory calculations in noncommutative gauge theories. More generally, deriving the early-time logarithmic response and the entanglement velocity \(v_E^{\mathrm{NC}}\) from the boundary theory would provide an independent test of their holographic interpretation and clarify the role of noncommutativity. Another interesting direction is to investigate whether the large-region dynamics admits a membrane-like description adapted to NCYM theory, following the approach of \cite{Mezei:2018jco}. Finally, studying the evolution of other boundary information measures, including those related to the EWCS, may help identify universal features of entanglement growth and the dynamics of mixed-state correlations in noncommutative field theories.

\subsection*{Acknowledgements}
We are very grateful to Mohammad Hasan Vahidinia for careful reading of the manuscript and his valuable comments. The work of M. Reza Mohammadi Mozaffar is supported by the Iran National Science Foundation (INSF) under project No. 40403689. Portions of the text were revised
with the assistance of ChatGPT (OpenAI) to improve English expression.

	\appendix
	\section{Consistency of the noncommutative Vaidya background}
	\label{app:nc-vaidya-consistency}
	
	We verify the sourced type IIB supergravity equations for the
	noncommutative Vaidya metric \eqref{eq:setup-Vaidya-metric}. The dilaton
	and form fields are kept at their static noncommutative D3-brane
	profiles, with the time dependence entering only through the blackening
	function in Eq.~\eqref{eq:setup-Vaidya-blackening}. We use the
	string-frame action and equations of motion in
	Eqs.~\eqref{eq:setup-IIB-action} and
	\eqref{eq:setup-Einstein-equation}.
	The signs of the modified Ramond--Ramond (RR)
	field strengths follow \cite{Sato:2002kv},
 while the general type IIB
	framework and the imposition of five-form self-duality after variation
	follow Refs.~\cite{Bergshoeff:1995as,Bergshoeff:2005ti}. Considering the action given by Eq.~\eqref{eq:setup-IIB-action}, we have
	the following field strengths in terms of the \(p\)-form potentials
	\begin{equation}
		H_3=dB_2,
		\qquad
		F_{\{1,3,5\}}=dC_{\{0,2,4\}},
		\qquad
		\widetilde F_{\{3,5\}}=F_{\{3,5\}}+C_{\{0,2\}}\wedge H_3.
		\label{eq:app-field-strengths}
	\end{equation}
	We work in the axion-free sector and impose self-duality only after
	varying the action.
	\begin{equation}
		C_0=0,
		\qquad
		F_1=0,
		\qquad
		\widetilde F_3=F_3,
		\qquad
		\star\widetilde F_5=\widetilde F_5.
		\label{eq:app-axion-free-self-duality}
	\end{equation}
	The dilaton profile is the one given in
	Sec.~\ref{subsec:nc-Vaidya-geometry}. The remaining potentials of the
	static noncommutative D3-brane solution may be chosen as
	\cite{Hashimoto:1999ut,Maldacena:1999mh}
	\begin{equation}
		B_2
		=
		\frac{R^2a^2}{z^4+a^4}\,
		dx_2\wedge dx_3,
		\;\;\;\;\;\;\;\;\;\;
		C_2
		=
		\frac{R^2a^2}{g_s z^4}\,
		dv\wedge dx_1.
		\label{eq:app-two-form-potentials}
	\end{equation}
	In each static region, the second expression is gauge-equivalent to the
	standard RR two-form in static coordinates. We retain the same radial profile, with no explicit \(v\)-dependence, across the Vaidya geometry. The 	corresponding three-form field strengths are
	\begin{equation}
		H_3
		=
		-\frac{4R^2a^2z^3}{(z^4+a^4)^2}\,
		dz\wedge dx_2\wedge dx_3,
		\;\;\;\;\;\;\;\;\;
		F_3
		=
		-\frac{4R^2a^2}{g_s z^5}\,
		dz\wedge dv\wedge dx_1.
		\label{eq:app-three-form-components}
	\end{equation}
	Let $\rho_5$ denote the volume form of the unit $S^5$ and choose the
	orientation
	$dv\wedge dz\wedge dx_1\wedge dx_2\wedge dx_3\wedge\rho_5>0$.
	The self-dual five-form is
	\begin{equation}
		\widetilde F_5
		=
		\frac{4R^4}{g_s}
		\left[
		\frac{h(z)}{z^5}\,
		dv\wedge dz\wedge dx_1\wedge dx_2\wedge dx_3
		+
		\rho_5
		\right].
		\label{eq:app-five-form}
	\end{equation}
	The Bianchi identities are satisfied directly
	\begin{equation}
		dH_3=0,
		\qquad
		dF_3=0,
		\qquad
		F_3\wedge H_3=0,
		\qquad
		d\widetilde F_5=0.
		\label{eq:app-form-bianchi-identities}
	\end{equation}
	With the conventions of Eq.~\eqref{eq:app-field-strengths}, the last
	identity and five-form self-duality also imply the five-form equation of
	motion. Moreover, the axion equation reduces to
	\begin{equation}
		H_{MNP}F_3^{MNP}=0.
		\label{eq:app-axion-equation}
	\end{equation}
	Every nonzero component of $H_3$ contains both $x_2$ and $x_3$, whereas
	every nonzero component of $F_3$ contains $x_1$. Raising the $v$ and $z$
	indices with the ingoing metric does not change this index mismatch.
	Therefore the contraction in Eq.~\eqref{eq:app-axion-equation}
	vanishes.
	
	The source in Eq.~\eqref{eq:setup-null-source} has vanishing trace,
	since it is proportional to $k_Mk_N$ with $k^Mk_M=0$. The trace of the
	Einstein equation may therefore be combined with the dilaton equation
	obtained from Eq.~\eqref{eq:setup-IIB-action}, giving
	\begin{equation}
		\nabla^2\Phi
		-2(\nabla\Phi)^2
		+\frac{1}{12}
		\left(
		H_3^2-e^{2\Phi}F_3^2
		\right)
		=0.
		\label{eq:app-dilaton-equation}
	\end{equation}
	The left-hand side vanishes identically using Eq. \eqref{eq:app-three-form-components} and the dilaton profile. In particular, no term proportional to $\partial_v m(v)$ is generated, so
	the Vaidya deformation requires no external dilaton source.
	For the axion-free field content in
	Eq.~\eqref{eq:app-axion-free-self-duality}, the RR and NS--NS three-form
	equations are
	\begin{equation}
		\begin{aligned}
			&\nabla_LF_3^{LMN}
			-\frac{1}{3!}H_{ABC}\widetilde F_5^{MNABC}
			=0,
			\\
			&\nabla_L\!\left(e^{-2\Phi}H_3^{LMN}\right)
			+\frac{1}{3!}F_{ABC}\widetilde F_5^{MNABC}
			=0.
		\end{aligned}
		\label{eq:app-three-form-equations}
	\end{equation}
	Up to antisymmetry, the only RR equation with nonzero terms is
	$(M,N)=(v,x_1)$. The two terms satisfy
	\begin{equation}
		\nabla_LF_3^{Lvx_1}
		=
		\frac{1}{3!}H_{ABC}\widetilde F_5^{vx_1ABC}
		=
		-\frac{16a^2z^4}{g_sR^4(z^4+a^4)}.
		\label{eq:app-RR-cancellation}
	\end{equation}
	For the NS--NS equation, the only equation with nonzero terms is
	$(M,N)=(x_2,x_3)$, for which
	\begin{equation}
		\nabla_L\!\left(e^{-2\Phi}H_3^{Lx_2x_3}\right)
		=
		-\frac{1}{3!}F_{ABC}\widetilde F_5^{x_2x_3ABC}
		=
		\frac{16a^2(z^4+a^4)}{g_s^2R^4z^4}.
		\label{eq:app-NS-cancellation}
	\end{equation}
	All remaining components vanish due to antisymmetry, the index structure of the three-forms, and the coordinate dependence of the background fields. Equations~\eqref{eq:app-RR-cancellation} and
	\eqref{eq:app-NS-cancellation} verify the two equations in
	Eq.~\eqref{eq:app-three-form-equations}; no external NS--NS or RR
	two-form current is required.
	
	For constant $m$, the fields introduced above reproduce the static
	noncommutative D3-brane solution
	\cite{Hashimoto:1999ut,Maldacena:1999mh}, and the source-free Einstein
	equation is satisfied componentwise. For a general mass profile
	$m(v)$, all terms that do not contain $\partial_v m(v)$
	cancel as in the static solution. The only new
	derivative-dependent contribution appears in the $(v,v)$ component.
	
	The contributions proportional to \(\partial_v m(v)\)
	in the Ricci tensor and dilaton Hessian are
	\begin{equation}
		\left.R_{vv}\right|_{\partial_v m}
		=
		\frac{z^3(3z^4-a^4)}
		{2(z^4+a^4)}
		\,\partial_v m(v),
		\qquad
		\left.2\nabla_v\nabla_v\Phi\right|_{\partial_v m}
		=
		\frac{2a^4z^3}
		{z^4+a^4}
		\,\partial_v m(v).
		\label{eq:app-vv-derivative-pieces}
	\end{equation}
	
	The form-field contributions are algebraic in $f(v,z)$ and contain
	no derivatives of the mass profile. Adding the two terms in
	Eq.~\eqref{eq:app-vv-derivative-pieces}, all explicit dependence on
	the noncommutative scale cancels, and one obtains
	$\frac{3}{2}z^3\partial_v m(v)$. This is precisely the $(v,v)$
	component of the external source defined in
	Eq.~\eqref{eq:setup-null-source}. All other
	$\partial_v m(v)$-dependent components vanish. The null energy condition is equivalent to $\partial_v m(v)\geq0$, which is satisfied by the smooth mass profile in Eq.~\eqref{eq:setup-smooth-profile}.
	With the normalization adopted in
	Eq.~\eqref{eq:setup-Einstein-equation}, the source in the trace-reversed
	string-frame equation obeys the dilaton-weighted identity
	\begin{equation}
		\nabla^M(e^{-2\Phi}T^{\mathrm{ext}}_{MN})	=
		0.
		\label{eq:app-weighted-source-conservation}
	\end{equation}
	This is the conservation law compatible with the sourced string-frame
	field equations.
	Thus, the metric \eqref{eq:setup-Vaidya-metric} with the static
	dilaton and form-field profiles satisfies the sourced  type IIB
	equations for arbitrary smooth $m(v)$, with the null source in
	Eq.~\eqref{eq:setup-null-source}. No external scalar or form-field
	current is generated, and the null energy condition requires a
	nondecreasing mass profile.

	\section{Details of the saturation regime}
	\label{app:saturation-details}
	
	In this appendix, we give the details of the saturation analysis in
	Sec.~\ref{subsec:saturation}. We work in the thin-shell limit and consider the monotonic
	shell-crossing branch described by
	Eqs.~\eqref{eq:setup-width-parametric}--
	\eqref{eq:setup-entropy-parametric}. In the parameter range considered
	in Sec.~\ref{subsec:static-finite-temperature}, the equilibrium surface
	is the deeper \(U_1\) branch selected by the minimum-area condition. We consider two limiting regimes: the near-horizon large-width expansion
	leading to Eq.~\eqref{eq:saturation-continuous-time}, and the
	fixed-\(z_b\) near-saturation expansion used in
	Eqs.~\eqref{eq:saturation-time-test} and
	\eqref{eq:saturation-quadratic}.
	
	We first consider the near-horizon large-width limit of
	Sec.~\ref{subsec:saturation}. The final blackening function has a simple
	zero, $f(z_h)=0$ with $f'(z_h)\neq0$, while the warp factor is finite and
	positive, $0<h(z_h)<\infty$. The endpoint of the time integral in
	Eq.~\eqref{eq:saturation-continuous-time-integral} therefore gives
	the second logarithm in Eq.~\eqref{eq:saturation-near-horizon}.
	For the width, write $y=z/z_b$ in the static integral
	$\ell/2=\mathcal F(z_b)$ and set $y=1-s$ near its upper endpoint. For
	$s,\eta\ll1$,
	\begin{equation}
		f(z_by)= -z_hf'(z_h)(s+\eta)+\mathcal O(s^2,s\eta,\eta^2),
		\;\;\;\;\;\;\;\;\;\;
		h(z_by) = h(z_h)+\mathcal O(s,\eta).
		\label{eq:app-saturation-local-horizon-expansions}
	\end{equation}
	The singular part of the width integral is
	\begin{equation}
		\left.\frac{\ell}{2}\right|_{\mathrm{sing}}
		=
		\sqrt{\frac{z_h}{6|f'(z_h)|h(z_h)}}
		\int_0^\Lambda\frac{ds}{\sqrt{s(s+\eta)}}=
		\sqrt{\frac{z_h}{6|f'(z_h)|h(z_h)}}\log\!\left(\frac{1}{\eta}\right)+\mathcal O(1),
		\label{eq:app-saturation-width-endpoint}
	\end{equation}
	where $\Lambda$ is fixed and small. Here $\mathcal O(1)$ denotes terms finite as $\eta\to0$; such terms may
	depend on $a$, $z_h$, $\delta$, and the fixed upper limit $\Lambda$.
	The above equation reproduces the
	first logarithm in Eq.~\eqref{eq:saturation-near-horizon}. Eliminating
	$\eta$ between the two logarithms gives
	Eq.~\eqref{eq:saturation-continuous-time}.
	
	We next use the parametrization introduced in
	Eq.~\eqref{eq:saturation-expansion}. The limit $\epsilon\to0$ is taken at
	fixed $z_b<z_h$ on a regular monotonic branch, and we assume
	$\mathcal F'(z_b)\neq0$. This condition is required for the
	fixed-width implicit-function expansion and fails at a turning
	point or cusp of the width--turning-point curve.
	The functions $\mathcal F$ and $\mathcal J$ are defined in
	Eq.~\eqref{eq:saturation-auxiliary-integrals}, while $H(z)$ is defined
	in Eq.~\eqref{eq:setup-radial-equation}. Using Eqs.~\eqref{eq:setup-vacuum-segment} and
	\eqref{eq:setup-shell-energy}, the black-brane integration constant and
	its expansion near $\epsilon=0$ are
	\begin{equation}
		\begin{aligned}
			E
			&=
			-\frac{g(z_c)}{2\sqrt{h(z_c)}}
			\left(\frac{z_t}{z_c}\right)^3
			\sqrt{1-\left(\frac{z_c}{z_t}\right)^6},
			&\qquad
			E
			&=
			-\frac{g(z_t)}{2\sqrt{h(z_t)}}\,
			\epsilon+\mathcal O(\epsilon^3).
		\end{aligned}
		\label{eq:app-saturation-energy}
	\end{equation}
	Further, adding and subtracting the static black-brane integrand in
	Eq.~\eqref{eq:setup-width-parametric}, we write $\ell/2=A_1-A_2+A_3+\mathcal F(z_t)$ where
	\begin{align}
		A_1
		&=
		\int_{z_c}^{z_t}
		\frac{dz}{\sqrt{h(z)}\sqrt{z_t^6/z^6-1}},
		\qquad
		A_2
		=
		\int_{z_c}^{z_t}\frac{dz}{\sqrt{H(z)}},
		\\
		A_3
		&=
		\int_{\delta}^{z_t}dz
		\left[
		\frac{1}{\sqrt{H(z)}}
		-
		\frac{1}{\sqrt{f(z)h(z)}\sqrt{z_t^6/z^6-1}}
		\right].
		\label{eq:app-saturation-width-decomposition}
	\end{align}
	Expanding near $z=z_t$ gives
	\begin{align}
		A_1=A_2
		=
		\frac{z_t}{3\sqrt{h(z_t)}}\,\epsilon
		+\mathcal O(\epsilon^3),
		\qquad
		A_3
		=
		-\frac{z_tg(z_t)}{6f(z_t)\sqrt{h(z_t)}}\,\epsilon
		+\mathcal O(\epsilon^2).
	\end{align}
	The linear terms in $A_1$ and $A_2$ cancel. At fixed width,
	$\ell/2=\mathcal F(z_b)$, while
	\begin{equation}
		\mathcal F(z_t)
		=
		\mathcal F(z_b)
		+\frac{z_b\sigma}{6}\mathcal F'(z_b)
		+\mathcal O(\epsilon^2).
		\label{eq:app-saturation-F-expansion}
	\end{equation}
	The fixed-width constraint then yields
	\begin{equation}
		\sigma
		=
		\frac{g(z_b)}
		{f(z_b)\sqrt{h(z_b)}\,\mathcal F'(z_b)}\,
		\epsilon
		+\mathcal O(\epsilon^2),
		\label{eq:app-saturation-sigma-epsilon}
	\end{equation}
	where the derivative is evaluated at fixed cutoff $\delta$. Subtracting Eq.~\eqref{eq:saturation-continuous-time-integral} from Eq.~\eqref{eq:setup-time-parametric}, we obtain $t-t_s^{(\mathrm{cont})}=B_1+B_2-B_3$ where
	\begin{align}
		B_1
		=
		\int_{z_b}^{z_c}\frac{dz}{f(z)},
		\qquad
		B_2
		=
		E\int_{\delta}^{z_t}
		\frac{h(z)\,dz}{f(z)\sqrt{H(z)}},
		\qquad
		B_3
		=
		E\int_{z_c}^{z_t}
		\frac{h(z)\,dz}{f(z)\sqrt{H(z)}}.
		\label{eq:app-saturation-time-decomposition}
	\end{align}
	To first order in $\epsilon$ we have
	\begin{align}
		B_1
		=
		\frac{z_b\sigma}{6f(z_b)}
		+\mathcal O(\epsilon^2),
		\qquad
		B_2
		=
		-\frac{g(z_b)}{2\sqrt{h(z_b)}}\,
		\mathcal J(z_b)\epsilon
		+\mathcal O(\epsilon^2),
		\qquad
		B_3
		=
		\mathcal O(\epsilon^2).
	\end{align}
	Using Eq.~\eqref{eq:app-saturation-sigma-epsilon} in these expressions
	reproduces Eq.~\eqref{eq:saturation-time-test}, including the coefficient
	$\omega$. Further, considering $S_{\mathrm{eq}}$ in
	Eq.~\eqref{eq:static-equilibrium-entropy} as a function of its turning
	point, we write the nearby dynamical entropy as $S(\ell,t)
	=
	C_1-C_2+C_3+S_{\mathrm{eq}}(z_t)$ with
	\begin{align}
		C_1
		&=
		2z_t^3\int_{z_c}^{z_t}
		\frac{dz}{z^6\sqrt{h(z)}\sqrt{z_t^6/z^6-1}},
		\qquad
		C_2
		=
		2z_t^3\int_{z_c}^{z_t}
		\frac{dz}{z^6\sqrt{H(z)}},
		\nonumber\\
		C_3
		&=
		2z_t^3\int_{\delta}^{z_t}\frac{dz}{z^6}
		\left[
		\frac{1}{\sqrt{H(z)}}
		-
		\frac{1}{\sqrt{f(z)h(z)}\sqrt{z_t^6/z^6-1}}
		\right].
		\label{eq:app-saturation-entropy-decomposition}
	\end{align}
	The corresponding leading terms are
	\begin{align}
		C_1=C_2=
		\frac{2\epsilon}{3z_t^2\sqrt{h(z_t)}}
		+\mathcal O(\epsilon^3),
		\qquad
		C_3
		=
		-\frac{g(z_t)}{3z_t^2f(z_t)\sqrt{h(z_t)}}\,
		\epsilon
		+\mathcal O(\epsilon^2).
	\end{align}
	The linear terms in $C_1$ and $C_2$ cancel. The remaining linear contribution cancels against the expansion of \(S_{\mathrm{eq}}(z_t)\), as follows from the relation between the static entropy and the strip width. The first integral
	\eqref{eq:setup-first-integral-J} implies
	$dS_{\mathrm{eq}}/d\ell=1/q^3$ along the static family, while
	$\ell=2\mathcal F(q)$. Hence,
	\begin{equation}
		q^3S_{\mathrm{eq}}'(q)
		=
		2\mathcal F'(q),
		\label{eq:app-saturation-HJ-identity}
	\end{equation}
	where both derivatives are evaluated at fixed $\delta$. Expanding
	$S_{\mathrm{eq}}(z_t)$ about $z_b$ and using
	Eq.~\eqref{eq:app-saturation-sigma-epsilon}, the coefficient linear in
	$\epsilon$ becomes
	\begin{equation}
		\begin{aligned}
			\left[S(\ell,t)-S_{\mathrm{eq}}(\ell)\right]_{\mathcal O(\epsilon)}
			={}&
			\frac{g(z_b)}{3z_b^2f(z_b)\sqrt{h(z_b)}}
			\left[
			\frac{z_b^3S_{\mathrm{eq}}'(z_b)}
			{2\mathcal F'(z_b)}
			-1
			\right]\epsilon
			={}&0.
		\end{aligned}
		\label{eq:app-saturation-linear-entropy-cancellation}
	\end{equation}
	Thus, the entropy difference starts at
	$\mathcal O(\!\epsilon^2)$. If the quadratic coefficient is nonzero and
	the physical continuous branch approaches equilibrium from below, the
	coefficient of the leading $\epsilon^2$ term in
	$S(\ell,t)-S_{\mathrm{eq}}(\ell)$ is negative. For $\omega\neq0$, this
	gives the quadratic time dependence in
	Eq.~\eqref{eq:saturation-quadratic}. This completes the fixed-cutoff derivation of
	the near-saturation time expansion and the cancellation
	of the linear entropy correction used in
	Sec.~\ref{subsec:saturation}.

\end{document}